\documentclass[11pt]{article} 
\usepackage{epsfig}
\usepackage{graphicx}
\usepackage{amsmath}
\usepackage{amssymb}	
\usepackage{amscd}	
\usepackage{caption2}
\usepackage{bbm}

\usepackage[thmmarks,amsmath,thref]{ntheorem}
\usepackage[all]{xy}

\theoremseparator{.}
\newtheorem{theor}{Theorem}
\newtheorem{prop}{Proposition}
\newtheorem{lemm}{Lemma}
\newtheorem{rem}{Remark}
\newtheorem{cond}{Conditions}
\newtheorem{cor}{Corollary }

\theoremstyle{nonumberplain}
\theoremheaderfont{\it}
\theoremseparator{:}
\theorembodyfont{\rm}
\theoremsymbol{$\Box$}
\newtheorem{prf}{Proof}
\newtheorem{prfs}{Scetch of proof of Theorem~\ref{thmainresult}}
\newtheorem{prftheorem2}{Proof of Theorem 2}
\usepackage{paralist}
\usepackage{stmaryrd}

\newcommand{\dbrack}[1]{\llbracket#1\rrbracket}
\newcommand{\bigbrack}[1]{\pmb{(}#1\pmb{)}}

\def\M{\mathcal{M}}
\def\H{\mathcal{H}}

\def\P{\mathcal{P}}
\def\I{\mathcal{I}}
\def\J{\mathcal{J}}
\def\U{\mathcal{U}}
\def\Ee{\mathcal{E}}
\def\C{\mathcal{C}}

\def\Lbf{\mathbf{\Lambda}}
\def\ka{\Bbbk}
\def\u{\mathbf{U}}
\def\w{\mathbf{w}}
\def\pii{\pmb{\pi}_I}
\def\x{\mathrm{x}}
\def\y{\mathrm{y}}

\def\KS{ {\scriptscriptstyle\mathrm{KS}}}
\def\BD{ {\scriptscriptstyle\mathrm{BD}}}

\def\D{ {\scriptscriptstyle\mathrm{D}}}
\def\L{ {\scriptscriptstyle\mathrm{L}}}
\def\nE{ n_{\scriptscriptstyle\mathrm{E}}}
\def\Op{\mathbf{Op}}
\def\Lip{\mathrm{Lip}}
\def\Level{\widetilde{I}}

\def\htower{l_{\scriptscriptstyle\mathrm{tow}}}
\def\falltower{\mathcal{D}}
\def\Pb{\mathbf{P}}

\begin{document}

\begin{center}
\vskip 0.5 cm
{\large \bf Entropic bounds on semiclassical measures for quantized
 one-dimensional maps
}

\vskip 0.5 cm

 BORIS GUTKIN
\vskip 0.3 cm

{\em Fachbereich Physik,
Universit{\"a}t Duisburg-Essen,\\ 
 Lotharstrasse 1,
47048 Duisburg, Germany}\\
{\small E-mail:  boris.gutkin@uni-due.de }

\end{center}
\vskip 1.0 cm

\begin{abstract}

\noindent
Quantum ergodicity asserts that almost all infinite sequences of 
eigenstates of a quantized ergodic system are equidistributed in the
phase space.  On the other hand, there are might exist exceptional
sequences which converge to different (non-Liouville) classical
invariant measures $\mu$.    By the remarkable result
of N. Anantharaman and S. Nonnenmacher math-ph/0610019, 
arXiv:0704.1564 (with H. Koch),  for Anosov geodesic flows  the metric entropy  of any
semiclassical measure $\mu$   must
be bounded from below.   The
result seems  to be optimal for uniformly expanding systems, but not in general case, where
it  might  become even  trivial  if the curvature of the Riemannian  manifold is strongly non-uniform. 
It has been conjectured by the same authors, that in fact,  a stronger bound (valid in general case) should hold.  

In the present work we consider such entropic bounds using the model of quantized
one-dimensional maps.  For a certain class of non-uniformly
expanding maps we prove Anantharaman-Nonnenmacher conjecture.
Furthermore, for these maps we are able to construct some  explicit
sequences of eigenstates which saturate the bound.  This demonstrates
that the conjectured bound is actually optimal in that case.

\end{abstract}

\vskip 2.0 cm
\section{Introduction}

The theory of quantum chaos concerns with the quantum systems whose  
 classical limit is  chaotic.  It is assumed in general,
 that chaotic dynamics induce certain characteristic patterns. For
 instance, the Random Matrix conjecture predicts that  statistical distribution
 of high-lying eigenvalues in a chaotic system   is the same as in
 certain ensembles of random matrices and depends only on symmetries of the
 system \cite{Bo}.  In the same spirit, it is believed that eigenstates of
 chaotic systems are   delocalized  over the whole available part of the phase
 space \cite{Be}, \cite{Vo} which is 
 totally different from  the case of integrable dynamics, where eigenstates
 are known to concentrate near KAM tori \cite{La}. 
The rigorous  implementation of that idea is known as {\it
 Quantum Ergodicity Theorem}.  
 It  was  first proven by A. I. Schnirelman for Laplacians on surfaces
 of negative curvature \cite{Sc} and later generalized \cite{Ze}, \cite{Co} and  extended to other systems e.g., ergodic
 billiards \cite{GL,ZZ}, quantized maps \cite{BDe} and general Hamiltonians \cite{HMR}.  

Very generally, the Quantum Ergodicity Theorem states that  for a
 classically ergodic system ``almost all'' eigenstates   in the semiclassical
 regime become uniformly   distributed over the phase space.
To give the    precise meaning of such a statement it is convenient
 to use the notion of measure. For a   Hamiltonian system a
 sequence of the  eigenstates $\{\psi_\ka, \,\,  \ka=1,\dots \infty\}$ generates the
 corresponding sequence of  the measures $\{d\mu_\ka=W_\ka(x,\xi)\,dxd\xi, \,\, \ka=1,\dots \infty\}$
 on the classical phase space, where the density $W_\ka(x,\xi)$ can be interpreted as the ``distribution'' of    $\psi_\ka$ over the phase space. Although the exact form of
 $W_\ka$ depends on the quantization procedure (e.g., Weyl, Anti-Wick
 quantization etc.), the limiting  {\it semiclassical measure}:
\begin{equation}\lim_{\ka\to\infty} \mu_\ka=\mu,\end{equation}
is invariant under the corresponding classical flow and does not depend on the choice of the quantization.  The Quantum
 Ergodicity theorem asserts that for ``almost all'' sequences of the
 eigenstates the limiting measure $\mu$ is actually the Liouville measure.

Since the Quantum
 Ergodicity theorem does not exclude possibility that  exceptional sequences of eigenstates produce non-Liouville classically invariant measures,  it makes sense to ask whether such measures might actually appear.
In the context of Anosov geodesic flows on surfaces of negative curvature it was  conjectured \cite{RS} that  a typical
 system  posses {\it ``Quantum Unique Ergodicity''} property, meaning that
 all sequences of  eigenstates converge to the   Liouville measure. 
However, there have been  only a
 limited number of rigorous results supporting this conjecture. So far, the most
 important one  was obtained by E. Lindenstrauss. In  
 \cite{Li} he  proved  that  all Hecke eigenstates of the Laplacian on compact
 arithmetic surfaces  are equidistributed. If (as widely believed) all
 the  Laplacian eigenstates  are  non-degenerate, this result would amount to the proof of
 Quantum  Unique Ergodicity for  the arithmetic  case.  On the other hand, it is known
 that exceptional sequences  actually do appear in some quantum
 systems. For quantum ``cat maps'' such sequences were identified in
 \cite{DeFN} \cite{FN}. The limiting measure  there could be, for instance,
 composed of two ergodic components:
 \begin{equation}
\mu=a\mu_{\L}+(1-a)\mu_{\D} , \qquad 1\geq a\geq 1/2, \label{nonmeasure}
\end{equation} 
where the first part $\mu_{\L}$ is the  Liouville measure
 equidistributed over the whole  phase space and the second part $\mu_{\D}$ is  the
 Dirac peak concentrated on a single unstable periodic orbit. Similar
  sequences of eigenstates have been also constructed  for the ``Walsh
 quantization`` of the baker's map  \cite{AN1}. For quantized
 hyperbolic automorphisms of higher-dimensional tori   there exists a different
 type of semiclassical measures which are Lebesgue measures on
 some invariant co-isotropic subspaces of the torus \cite{Kel}. 

As we know that non-Liouville semiclassical measures  do appear (at least) in some systems,
it would be of great interest to understand which kind of them   might exist  in a general case.
Quite recently, it  has been  proven by  N. Anantharaman and S. Nonnenmacher  \cite{An}, \cite{AN2}, \cite{AKN} (with H. Koch)  that for the  Laplacian on a compact Riemannian  manifold with 
 Anosov geodesic flow the
metric (Kolmogorov-Sinai) entropy $H_{\KS}(\mu)$ of any semiclassical
 measure $\mu$ must satisfy  certain bound. Particularly, in 
the two-dimensional case   the following result holds \cite{AKN}:
\begin{equation}H_{\KS}(\mu)\geq\int|\log
 J^u(x)|d\mu-\frac{1}{2}\lambda_{\max}, \label{Stheor}\end{equation} 
 where  $J^u(x)$ is the unstable Jacobian of the flow at the point $x$
 and  $\lambda_{\max}$ is the 
maximum expansion rate of the flow. If the maximum  expansion rate is
 close to its average value, this remarkable bound gives a valuable  information on
 $\mu$ itself. In particular, for surfaces  with a constant
 negative curvature  this remarkable  bound implies that maximum ``half'' of the
  measure might  concentrate on periodic orbits. On the other hand, if the
 expansion rate  varies a lot,   the above bound  does not give any
 information, as the right hand side of (\ref{Stheor}) becomes negative. Thus,
it is natural to expect that (\ref{Stheor}) is not an optimal result, and a 
stronger bound might exist in a general case. Such a bound has been conjectured in \cite{An, AN1}. It states that for chaotic systems  a semiclassical measure must satisfy:
 \begin{equation} \label{Sconj} H_{\KS}(\mu)\geq\frac{1}{2}\int|\log
 J^u(x)|d\mu.
\end{equation}
Assuming that  the conjecture is true, it provides  a restriction on  the class of possible semiclassical measures in
 general case. In particular, for semiclassical measures of the type  (\ref{nonmeasure}) the bound (\ref{Sconj}) would imply that  Liouville part should be always present and its proportion satisfy $a\geq\frac{\lambda_\D}{\lambda_{\mathrm{av}}+\lambda_\D}$, where 
$\lambda_{\mathrm{av}}$ is the average Laypunov exponent (with respect to the Liouville measure) and $\lambda_\D$ is the Laypunov exponent for the periodic orbit where  $\mu_{\D}$ is localized.


\section{Model and  statement of the main results}

 The central  purpose of this paper is to provide  
 support for the conjectured bound (\ref{Sconj}) using the model  of
  quantized
   one-dimensional piecewise linear maps.  A procedure for quantization of
   one-dimensional linear maps was originally  
  introduced in \cite{PZK} in order   to
generate      families of quantum graphs with some special properties.   Being much simpler on the technical 
 level,  these models still exhibit  characteristic
 properties of   typical quantum chaotic Hamiltonian systems. Most importantly,  it turns out that the  quantum evolution here follows the classical evolution till the  (Ehrenfest) time which grows
logarithmically with  the dimension of the Hilbert space.\footnote{As we deal in the present paper with a discreate time evolution, the term "time" stands here and after for the number of  iterations of  either classical or quantum maps.}  Note also that, as will be shown in the body of the paper,   the construction is  closely related to the  Walsh quantized baker's maps in \cite{AN1}.

In the
present work we will consider Lebesgue measure preserving  maps $T:[0,1]\to [0,1]=:I$ consisting of
several linear branches. More specifically,  let   $\{ I_j, \, \, j=1,\dots l\}$ be a partition of the unite interval $I=\cup^l_{j=1}I_j$  into  $l$ subintervals. 
At each subinterval $I_j$,  $T$ is then defined as
a simple  linear 
map $T:I_j\to I$:
\begin{equation}T(x)= x\Lambda_{j}+b_j, \qquad \mbox{
for } x\in I_j,
 \,\,\, j=1,\dots l.\label{mapgeneral}\end{equation}

\begin{cond} {\rm We consider maps $T$  of the form  (\ref{mapgeneral}) 
  satisfying the following conditions:} 
\end{cond}
 \begin{itemize}
\item  $\sum^l_{j=1}\Lambda^{-1}_j=1$ and  $\Lambda_i$, $i=1,\cdots l$ are
 integers larger then one.
\item   Each subinterval $I_j$ is mapped by $T$ upon  the whole  unite interval
 $I$.  Correspondingly,  the Lebesgue measure of each  $I_j$  equals to   $\Lambda^{-1}_{j}$ and $b_1=0$, $b_i=-\Lambda_i(\sum_{k<i}\Lambda_k^{-1})$ for $1<i\leq l$.

\end{itemize}
\begin{rem} 
{\rm The first  condition above is essential.  It implies that the map is Lebesgue measure preserving, chaotic
 and the set of endpoints of partitions $\M_{\ka}$ is forward invariant under the
 action of $T$ (see below).
The second  condition  is imposed solely for the sake of simplicity of
 exposition. It implies that $i$'s branch of $T$ ``starts'' from a point $x_i$, where $T(x_i)=0$  and ``ends'' at the point $x_{i+1}$, where $T(x_{i+1})=1$.
In principle, most of the results of the paper can be extended to a more
 general  class of expanding piecewise linear maps considered in \cite{PZK}.  }
\end{rem}

We will now briefly describe the  procedure introduced by P.~Pako{\'n}ski {\it et al} \cite{PZK} for quantization of such maps.
 Let
 $\M=\{E_i,\,\, i=0,\dots N-1\}$ be  the  partition of  $I$ into   $N$ intervals
 $E_i=[i/N,(i+1)/N]$, $i=0,\cdots N-1$ of  equal  lengths.
 For the interval $E_i$ we will denote by $\beta_{+}(E_i)$
 ($\beta_{-}(E_i)$)  right (resp. left) endpoint of $E_i$ and by
 $\beta(\M)=\cup_{i=1}^{N}\beta_{\pm}(E_i)$ the set of all endpoints of the partition $\M$.
 Obviously both   $\M$ and  $\beta(\M)$ are uniquely
 determined by the size $N$ of the partition.  
In what follows we will consider  an increasingly refined  sequence of
 the above partitions  $\M_\ka$  with the sizes $N_{\ka}$,  $\ka=1,\dots \infty$.

\begin{cond} {\rm Given a map $T$ satisfying Conditions 1 we impose the
 following conditions on the sequence of $\M_\ka$:} 
\end{cond} 
 \begin{itemize}
\item Each partition $\M_\ka$  is a refinement of the previous one. That
  means  for each $\ka\geq 1$,   $N_{\ka+1}/N_{\ka}$  is  an integer number
 greater then one.

\item The set of the endpoints of the initial  partition $\M_1$ must include
 all  singular points  of $T$ i.e.,  $\beta(\M_1)\supseteq \beta(I_i)$
 for all $i=1,\dots l$. 
\end{itemize}
For a map  $T$ satisfying Conditions 1  and a sequence of  partitions $\M_\ka$, $\ka=1,\dots\infty$ satisfying Conditions 2  consider the sequence of the corresponding transfer (Frobenius-Perron) operators given by  $N_\ka\times N_\ka$ doubly stochastic matrices $B_\ka$, whose elements read as:
\begin{equation} B_\ka (i,j)=\frac{|E_i\cap T^{-1}E_j|}{|E_i|}=\left\{\begin{array}{lc} \Lambda_i^{-1} &\mbox{  if  } E_i\cap T^{-1} E_j\neq \emptyset\\
0 & \mbox{  otherwise.  } \end{array}\right.\label{bmatrix}\end{equation} 
 We will call a
piecewise linear
 map $T$ {\it quantizable} if   there exists a sequence of
partitions $\M_\ka$, $\ka=1,\dots\infty$ such that for each  matrix 
$B_\ka$ one can find  a  unitary matrix  $U_\ka$ of the same dimension 
satisfying
\begin{equation} B_\ka (j,i)= |U_\ka (i,j)|^2.\label{umatrix}\end{equation} 
  for each matrix
element $(j,i)$;  $j,i\in\{1,\dots N_\ka\}$.\footnote{
Note that our definition for 
$U$ matrix corresponds to the adjoint of the
corresponding quantum evolution in \cite{PZK}, \cite{BKS}.}
For quantizable
 maps the matrices $U_\ka$ are regarded as
``quantizations''
 of  $B_\ka$  and  play the role of  quantum evolution
operators
  acting on
  $N_\ka$-dimensional Hilbert space
$\H_\ka\simeq\mathbbm{C}^{N_\ka}$. 
As an example,  consider the following linear map (see fig. 1a):
\begin{equation}T(x)=2x\mod 1,\qquad x\in
 [0,1].\label{map}\end{equation}
\begin{figure}[htb]
\begin{center}
\includegraphics[height=3.5cm]{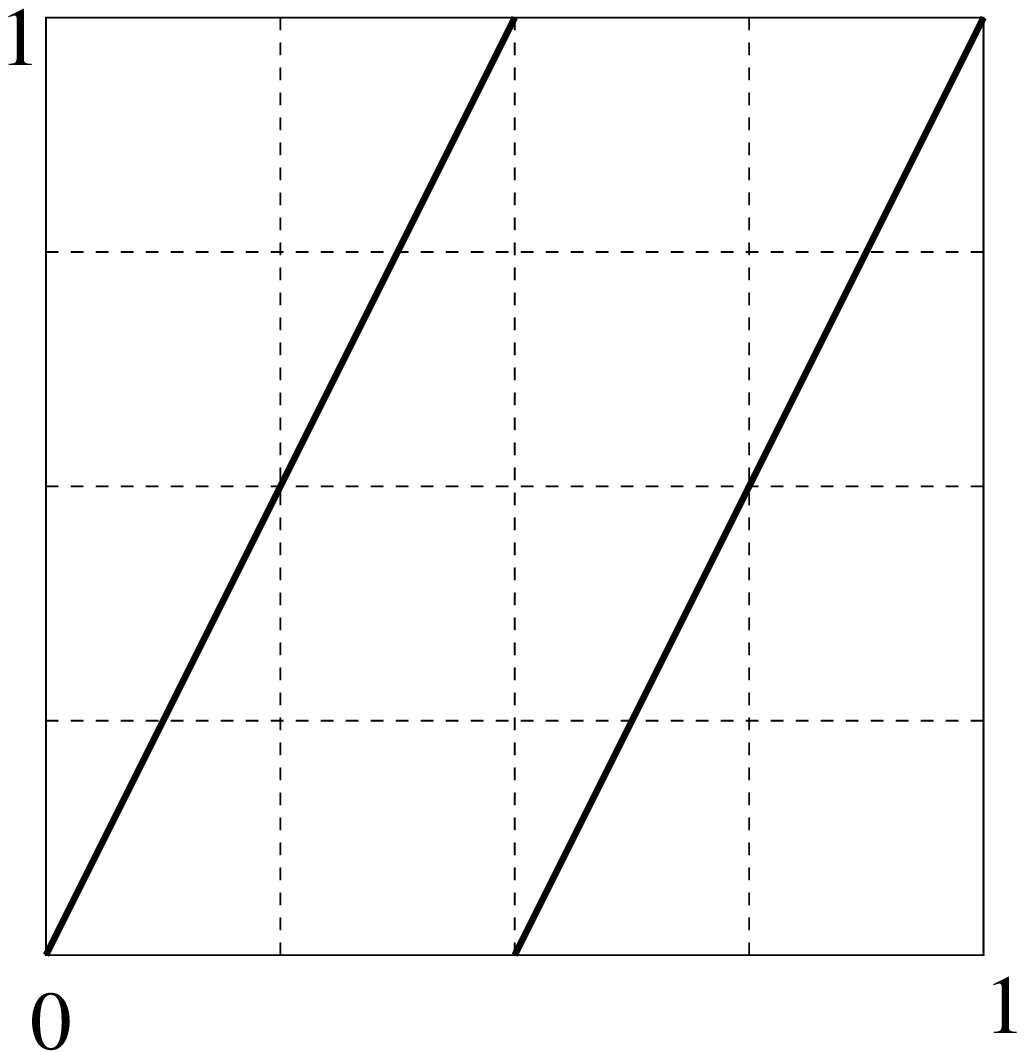}
\hskip 2.1cm
\includegraphics[height=3.5cm]{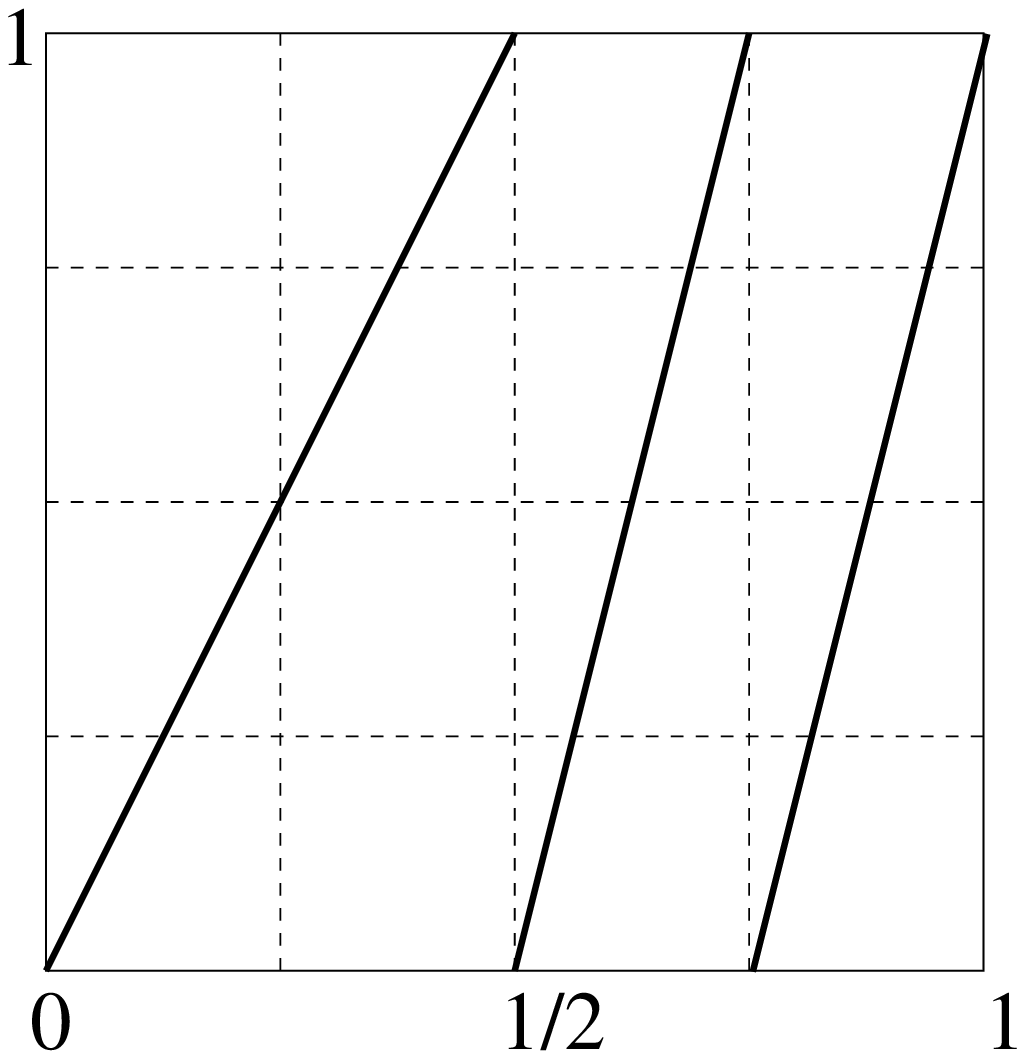}
\end{center}
\caption{\small{Linear  maps with uniform (left) and non-uniform slopes
 (right) which allow ``tensorial'' quantization.  }}
\end{figure}
 Here for the
sequence of
 partitions  $\M_\ka$  of the unite interval into
$N_\ka=2^\ka$ equal
 pieces, the matrix elements $B_\ka(i,j)$ of the
  classical transfer operators take the  values $1/2$ if $j=2i$, $j=2i-1$, $j+N_\ka=2i$, $j+N_\ka=2i-1$ and
 $0$ otherwise:
\begin{eqnarray*} 
B_2=\frac{1}{2}\left(\begin{array}{cc}
1 & 1 \\ 
1 & 1\\
 \end{array} \right), 
\qquad
B_4=\frac{1}{2}\left(\begin{array}{cccc}
1 & 1 & 0 &0\\ 
0 & 0 & 1 & 1\\
1 & 1 & 0 &0\\ 
0 & 0 & 1 & 1\\
 \end{array} \right), \qquad\cdots
\end{eqnarray*}
  Note that the structure of $B_\ka$, actually,  resembles the structure of the map $T$ (rotated clockwise  by $\pi/2$).
It is easy to see that the map  (\ref{map}) is quantizable. By a permutation of rows $B_\ka$ can be brought into the block diagonal form such that every block is $2\times 2$ matrix $B_2$ whose all elements are $1/2$. Thus the question of the quantization of $T$ reduces to finding of a unitary $2\times 2$ matrix $\u$ satisfying $|\u(l,m)|^2=1/2$ for all its elements. 
  The appropriate choice is given, for instance, by  the discrete Fourier
 transform: $\u (l,m)=\frac{1}{\sqrt{2}}\exp{(\pi i
 lm)}$. This example can be straightforwardly generalized to all other maps with a uniform slope. 
The question of  the quantizability of general piecewise
linear maps will be  discussed in the body of the paper.

 Note that the above quantization of  one-dimensional
piecewise linear
 maps is just a  formal procedure for generation of
unitary matrices
 $U_\ka$. 
To turn it to  a  ``meaningful'' quantization one
needs, in
 addition, to make a connection between  classical observables on the
unite
 interval and  the corresponding quantum  observables
  on the Hilbert
 space $\H_\ka$. Such a  quantization procedure has
been 
 introduced in \cite{BKS}.  With a classical
observable $f\in
 L^2[0,1]$
 one associates the sequence  of the quantum observables
$\Op_{\ka}(f)$,
 defined by the diagonal matrices of the dimension
$N_{\ka}$  whose
 components $\Op_{\ka}(f)_{j,j}$ equal to the average
value of $f$ at
 $j$'s
 element  of the partition $\M_\ka$. The key
observation   making
 the
 above  quantization interesting   is the existence
of  the
 semiclassical  correspondence ({\it Egorov property})
between evolutions
 of 
 classical and quantum observables. Precisely, for a
Lipschitz
 continues
 observable $f(x)$ one has \cite{BKS}:
\begin{equation}||U^*_\ka\Op_{\ka}(f)U_\ka-\Op_{\ka}(f\circ
T)||=O\left(\frac{1}{N_{\ka}}\right).\label{berkergodicity}\end{equation}
Note that,  the size of the partition  $N^{-1}_\ka$
plays here the role
 of the Planck constant and the semiclassical limit
corresponds to
 $\ka\to \infty$.

Equipped with the above quantization procedure we can
 define now the
 sequence of the semiclassical measures associated with  the 
eigenstates  of
 $U_\ka$. For  $ \psi_{\ka}\in\H_\ka$, $
U_\ka\psi_\ka=e^{i\theta_\ka}
 \psi_\ka$, $\ka=1,\cdots \infty$  we define 
$\mu_\ka$ through  the
 relationship:
\begin{equation} \int_{I}f(x)\, d\mu_\ka(x)=\langle
 \psi_\ka\Op_\ka(f)\psi_\ka\rangle.
\label{rietz}\end{equation}
We will be concerned with the possible semiclassical
limits   of
 $\mu_{\ka}$ as $\ka\to\infty$ and call any 
such limiting
 measure $\mu$  as  semiclassical measure. 
Speaking informally $\mu$
 characterizes the
 possible sets of the localization on the interval
$[0,1]$ of  the
 eigenstates of quantized maps. 
 (An alternative point of view (see  \cite{BKS}) is to
look at
 such limits as ``scars'' on the sequence of quantum
graphs defined by
  $U_\ka$.)
An immediate  consequence of the Egorov property is
that any
  semiclassical measure 
 $\mu$ must be invariant under
  the map $T$. Indeed, since  $\psi_\ka$ is an
eigenstate of
 $U_\ka$:
\begin{equation}
\int_{I}f(x)\, d\mu_\ka(x)=\langle \psi_\ka\,
 U_\ka^{*}\Op_\ka(f)U_\ka\psi_\ka\rangle 
=\int_{I}f(T(x))\, d\mu_\ka(x)
+O\left(\frac{1}{N_\ka}\right),
\end{equation}
and the  invariance of $\mu$ follows immediately after taking  the limit $\ka\to\infty$. As
 there exist many  classical measures preserved by
$T$, the invariance
 alone
 does not determine  all  possible outcomes for the
semiclassical measures.
 Similarly to  Hamiltonian systems,  using
Egorov property 
one can   show  by standard methods (see e.g.,
\cite{Db})
 that almost any sequence of the eigenstates gives rise to
 the Lebesgue
 measure in  the semiclassical limit (this  was
proved in
 \cite{BKS} by somewhat a different method).

\begin{theor} {\rm (Quantum Ergodicity
\cite[Thm.~2]{BKS}.)}  Let $T$ be a quantizable map (\ref{mapgeneral}) satisfying Condition~1 and let $U_\ka$, $\ka=1,\dots\infty$ be a sequence of its quantizations 
 with eigenstates $\psi^{(i)}_\ka$, $i=1,\dots N_\ka$. Then for each $\ka$ there exists  subsequence of $\mathcal{N}_\ka$  eigenstates: $\pmb{\Psi}_\ka:=\{\psi^{(i_1)}_\ka,\dots \psi^{(i_{\mathcal{N}_\ka})}_\ka\}$ such that $\lim_{\ka\to\infty}\mathcal{N}_\ka/ N_\ka=1$ and  for any  sequence of
 eigenstates  $\psi_{\ka_j}\in\pmb{\Psi}_{\ka_j}$, $j=1,\dots\infty$  and a Lipschitz continues function $f$ one has: 
\begin{equation}
\lim_{j\to\infty}\langle\psi_{\ka_j}
\Op_{\ka_j}(f)\psi_{\ka_j}\rangle
 =
 \int_{I}f(x)\, dx.\end{equation} 
\label{quantumergodicity}
\end{theor}

In the present  paper we go  beyond the Quantum
Ergodicity and
 ask about  the possible exceptional semiclassical
measures. Our first
 result is the precise analog of the bound
(\ref{Stheor}):
\begin{theor} Let $T$ be a quantizable piecewise
linear map 
 (\ref{mapgeneral}) satisfying Condition~1. Let $U_\ka$,  $\ka=1,\dots\infty$  be  a sequence of its quantizations  and  let $\psi_\ka$, $\ka=1,\dots\infty$  be some subsequence of its eigenstates.  Then the following bound holds for the metric entropy of the corresponding  semiclassical measure $\mu$:  
\begin{equation}H_{\KS}(T,\mu)\geq\int_I\log\Lambda(x)\,d\mu(x)-\frac{1}{2}\log\Lambda_{\max}
=\sum^l_{j=1}\mu(I_j)\log\left(\Lambda_{j}\right)
-\frac{1}{2}\log\Lambda_{\max},
 \label{firstresult}
\end{equation}
where $\Lambda_{\max}:=\max_{1\leq j\leq l}\Lambda_{j}$ and
$\mu(I_j)$ are the
 measures of the intervals $I_j$.
\label{thfirstresult}
 \end{theor}
As it is clear, that this bound is not  optimal  for
the  maps
 with non-uniform slopes, one would like
to have a stronger
 result, analogous to the conjectured one
(\ref{Sconj}).  In the present  we are able to prove such a bound  for a
particular subclass of
 piecewise linear maps (\ref{mapgeneral}). Namely, in the body of the paper we show that   the maps $T_p$ whose slopes
are given by the
 powers of the same integer number $p$ (see fig. 1b for
an example of such a
 map), allow  a special type of ``tensorial'' quantizations. For maps $T_p$ quantized in that way 
we prove the analog of
Anantharaman-Nonnenmacher  conjecture.
\begin{theor} Let $T_p$ be a  map of the form:  $T_p
 (x)=\Lambda_j x\mod{1}$, $ \Lambda_j=p^{n_j}$ for $x\in I_j, j=1,2\dots l$ and let $U_\ka$, $\ka=1,\dots\infty$ be  a sequence of ``tensorial'' quantization of $T_p$.  Then  for any sequence of  eigenstates  $\psi_\ka$ of $U_\ka$,
 $\ka=1,\dots\infty$  the corresponding  semiclassical measure $\mu$ satisfies:
\begin{equation}H_{\KS}(T_p,\mu)\geq
\frac{1}{2}\sum^l_{j=1}\mu(I_j)\log\left(\Lambda_{j}\right).\label{mainresult}
\end{equation}
\label{thmainresult}\end{theor}
Furthermore, for these 
 maps there exists  an explicit 
  construction of certain sequences of eigenstates of $U_\ka$. Using
 these
 eigenstates we obtain  a set  of  semiclassical
measures which can be
 subsequently
 analyzed to test  (\ref{mainresult}). It turns out
that some of these semiclassical measures, in
fact,   saturate 
  the bound   implying that the
result is 
 sharp.  

The paper is organized as follows. In Section 3 we
deal with  a general
 construction of unitary  evolutions for  piecewise-linear maps and
prove ``quantizability``  for   a wide class of  maps satisfying Conditions~1. Here  we also  introduce  a special
class of tensorial
 quantizations  for the maps $T_p$ whose slopes are given by the powers of an integer $p$.  In Section 4 we review the
construction in
 \cite{BKS} for   quantization of observables and prove the Egorov
 property up to the Ehrenfest time.  In Section 5 we connect metric
 entropy for the semiclassical measures with certain type of quantum  observables.
 Based on the method of \cite{AN2} we then prove Theorem~\ref{thfirstresult}
in Section 6  using the Entropic Uncertainty Principle.
Section 7 is devoted to the proof of Theorem~\ref{thmainresult}. Finally,
 in Section 8 we explicitly construct certain class of semiclassical measures for
tensorial  quantizations  of maps $T_p$ and test the bound (\ref{mainresult}). The concluding remarks are presented in Section 9.


\section{Quantization of one-dimensional piecewise linear maps  }

We will consider now in more details the quantizations of  Lebesgue measure preserving
  piecewise linear maps $T$ of the form (\ref{mapgeneral}).
Note that each map satisfying Condition 1 is uniquely
 determined by the ordered set of its slopes $\Lbf=\{\Lambda_1,\dots
 \Lambda_l\}$, so the notation $T=T_{\Lbf}$ will be often used to define the
 corresponding  map. 
Recall that a piecewise linear map $T_{\Lbf}$  is "quantizable" if there exists an infinite sequence of partitions $\M_\ka$ of unite interval $I$ such that the  corresponding evolution matrices   
$B_\ka$ allow representation (\ref{umatrix}).  In general, it is  a  non-trivial problem to determine whether  a  doubly stochastic matrix has such a representation in terms of a unitary matrix (see \cite{PZK}, \cite{ZKSS} and references there).  So, in principle, it is not clear in advance which of the  maps $T_\Lbf$  are actually   ``quantizable''.  It is our purpose here to show   that the  class of quantizable   piecewise linear maps is wide and contains many interesting maps.

\subsection{General quantization}

  As has been already mentioned a map  with a uniform slope is  quantizable by means of the discrete Fourier transforms. Hence,  a non-trivial question is about ``quantizability'' of the  maps $T_\Lbf$,     $\Lbf=\{\Lambda_1,\dots\Lambda_l\}$ with at least two different $\Lambda_i$. Let $\Lambda_{i_1},\dots\Lambda_{i_\ell}$, $\ell>1$ be the maximal set of  different slopes in $\Lbf$, i.e., $\Lambda_{i_n}\neq\Lambda_{i_m}$ for $n\neq m$.
Assuming that each slope  $\Lambda_{i_k}$ has a multiplicity $m_k\geq 1$, the Lebesgue measure preservation condition
\begin{equation}\sum_{k=1}^{\ell}\frac{m_k}{\Lambda_{i_k}}=1,\label{lebeasque} \end{equation}
 imposes certain restrictions on the values of $\Lambda_{i_k}$, $m_k$.  In particular, it is clear that the set  $\Lambda_{i_k},\,\, k=1,\dots \ell$ must have a greatest common divisor  $p$ large then one. This means
\[ \Lambda_{i_k}=p\bar{\Lambda}_k\qquad \bar{\Lambda}_k \in\mathbbm{N} \mbox{ for } k\in\{1,\dots \ell\}.\]
Assume now that all the numbers $ \bar{\Lambda}_i $ are relatively prime. Then it follows  immediately from (\ref{lebeasque}) that  $m_k$'s  are of the form $m_k=\bar{m}_k\bar{\Lambda}_k $, $ \bar{m}_k \in\mathbbm{N}$, $k\in\{1,\dots l\}$, where  $\sum_{k=1}^l  \bar{m}_k=p$.  We are going  now to show   that the maps  $T_\Lbf$ whose  slopes satisfy the above conditions  are quantizable. 
\begin{theor} Let $T_{\Lbf}$ be a map  satisfying Condition 1  with the  slopes $ \Lambda_i=p\bar{\Lambda}_i$, $ \bar{\Lambda}_{i+1}\geq \bar{\Lambda}_i$ of multiplicities $m_i$,   $i\in\{1,\dots l\}$  such that  $p\in\mathbbm{N}$ and $\bar{\Lambda}_i$'s are relatively prime integers, then  $T_{\Lbf}$ is ``quantizable''. \label{quantisability}
\end{theor}
\begin{prf} 
As the first step  notice that $ T_\Lbf$  can be represented as the composition of  the uniformly expanding map $\bar{T}_{p}$ and the "block diagonal"  map $T_{\BD}$,  whose slopes are uniform at each block.
\begin{lemm} Let  $T_{\Lbf}$  be  a map as defined above,  then  $ T_\Lbf=\bar{T}_{p}\circ T_{\BD}$, where $T_{p}(x)=xp\mod 1$ and  
\[T_{\BD}(x)= \left(\Lambda_i x \mod 1 \right)/p+b_i,  \qquad \mathrm{ for } \,\, x \in [ b_i, b_{i+1}],\,\,\qquad  b_i=\sum_{j<i}\frac{m_j}{\Lambda_j}, \]
where $m_i$ is the multiplicity of $\bar{\Lambda}_i $. 
\end{lemm} 
\begin{prf} Straightforward calculation. \end{prf}
The parameters entering into the definition  of $T_{\BD}$ have the following simple meaning. The points   $b_i$, $b_{i+1}$  mark the position of $i$'s block which is  the square of the  size $\frac{m_j}{\Lambda_j}$.  Inside  of each such block the map  $T_{\BD}$  acts  as a  piecewise linear map with the  uniform expansion rate $\bar{\Lambda}_i$.  \\

\noindent {\bf Example.}   To illustrate the above lemma consider as an example  the  map  with the slopes $6$ and  $4$: 
\begin{equation} T(x)=\left\{\begin{array}{ll} 6x\!\! \mod 1&\mbox{  if  }x\in[0,1/2) \\
4x \!\!\mod 1& \mbox{  if  }x\in[1/2,1] .\end{array}\right.\label{map4}\end{equation}
  As shown  in fig.~2,  $T$ can be decomposed into  the  uniformly expanding map $\bar{T}_{2}=2x\!\!\mod 1$  and  the "block diagonal" map: 
\begin{equation*} T_{\BD}(x)=\left\{\begin{array}{ll} (6x \!\!\mod 1)/2 &\mbox{  if  }x\in[0,1/2) \\
(4x \!\!\mod 1)/2 +1/2& \mbox{  if  }x\in[1/2,1] .\end{array}\right.
\end{equation*}
\begin{figure}[htb]
\begin{center}
\includegraphics[height=3.5cm]{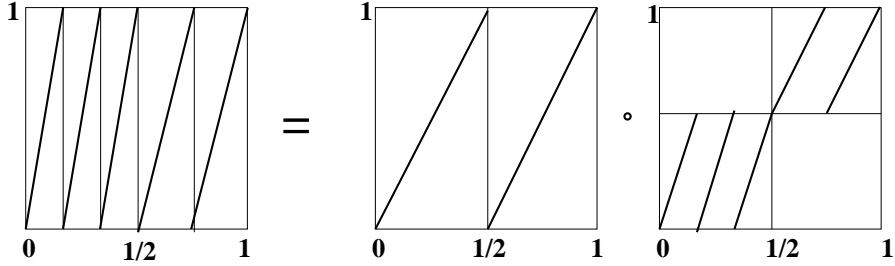}
\end{center}
\caption{\small{ A ``generic'' map (\ref{map4}) and its decomposition into   the uniformly expanding   and the  "block diagonal"  parts. }}
\end{figure}

Let us  now define  a set of partitions $\M_\ka$  of $I$ by  setting their seizes   $N_\ka$.  Take $N_0=p\prod_{i=1}^l\bar{\Lambda}_i$, then   $N_\ka=(N_0)^\ka$ for $\ka=\{0,\dots \infty\}$.  It is  clear that these partitions satisfy Conditions 2.  For each partition  $\M_\ka$  denote    by $\bar{B}_\ka$,  $B_\ka^{\BD}$ the corresponding evolution  operators  for  the map   $\bar{T}_{p}$  and $T_{\BD}$  respectively.  Note that both  $\bar{B}_\ka$ and  $B_\ka^{\BD}$ are quantizable i.e., one can find  unitary matrices $\bar{U}_\ka$,  $U_\ka^{\BD}$  satisfying (\ref{umatrix}). 
Indeed,  this is completely  obvious for $\bar{B}_\ka$   as   $\bar{T}_{p}$ has the uniform slope.  Since  $B_\ka^{\BD}$  has  the block diagonal form, the   corresponding quantum evolution $U_\ka^{\BD}$ can be defined  as the block diagonal matrix of the same structure where  each block is quantized with the help of   the discrete Fourier transform.   Given  matrices   $\bar{B}_\ka$,  $B_\ka^{\BD}$, and the quantizations   $\bar{U}_\ka$,  $U_\ka^{\BD}$
one can easily construct the transfer operator for the composition map   $ T_\Lambda=\bar{T}_{p}\circ T_{\BD}$ and the corresponding quantization. 
\begin{lemm} Let  $T_{\Lbf}$,  $\M_\ka$ be  the map  and partition as above  and let $B_\ka$ be  the corresponding evolution operator,  then  $B_\ka=B_\ka^{\BD}\bar{B}_\ka $  and the matrix  $U_\ka=\bar{U}_\ka U_\ka^{\BD}$ satisfies  (\ref{umatrix}). 
\end{lemm} 
\begin{prf} Straightforward check. \end{prf}
From this  the proof of the theorem follows immediately.
\end{prf}

\subsection{"Tensorial" quantizations}
 In  this subsection we will consider a special class of the  maps $T_\Lbf$, for which all $\Lambda_i=p^{n_i}$ are powers of some integer $p$. We will denote such maps by $T_p$. These maps are of interest as they posses  several  peculiar properties. In particular, as we show  below, $T_p$ allow a special type of ``tensorial'' quantizations which will be of use in the subsequent  parts of the paper.\\

\noindent{\bf Maps with a uniform slope.} We will first  consider  piecewise linear
  maps with the uniform 
 slope $\Lambda_j\equiv p\in\mathbbm{N}$ i.e, the maps: 
\begin{equation} \bar{T}_p( x )=p x \!\!\mod{1}, \qquad  x \in I \label{uniformmap}.\end{equation}
 (Here and after we will use the bar symbol  to distinguish the above  uniform maps from non-uniform ones.) For any point $x\in I$ it will be  convenient to use p-base numeral system: $x=0.\x_1\x_2\x_3\dots$,    $\x_i\in\{0,\dots,p-1\}$ to represent  $x$.  Obviously, each point   is then encoded by an infinite  sequence (not necessarily unique) of symbols $\x_1,\x_2,\x_3\dots$.
With such representation for the points in $I$ the action of $\bar{T}_p$ becomes equivalent to the simple shift map:
\begin{equation}\bar{T}_p:\x_1\x_2\x_3\x_4\dots\to \x_2\x_3 \x_4 \x_5\dots.\label{schiftmap}\end{equation}

In the following we will use  symbol $\x=\x_1\x_2\x_3\dots \x_m$ for  both finite and infinite sequences with the  notation $|\x|:=m$ reserved for the length of the sequence. So for $\x$ with  $|\x|=\infty$ the symbol  $\x$ will stand for the corresponding point $x=0.\x$  in the interval $I$. 
For a sequence $\x$, with finite $|\x|=m$  we will use notation  $\dbrack{\x}$  to  denote  the corresponding cylinder set, where the point $x\in\dbrack{\x}$ if the first $m$ digits of $x$ after the point coincide with  $\x_1,\x_2,\dots \x_m$. 
 For any  map  $\bar{T}_p$, there
 exists a sequence of natural Markov partitions $\M_\ka$ into $N_\ka=p^\ka$ cylinder sets of the length $\ka$:  
\[\{ E_\x=\dbrack{\x}, |\x|=\ka \}.\] 
The corresponding transfer operator  is then given by the matrix $B_\ka$, whose matrix elements: 

\begin{equation} B_\ka (\x,\x')=\left\{\begin{array}{lc} p^{-1} &\mbox{  if  } \x_i=\x'_{i+1}, \qquad i=1,\dots \ka-1\\
0 & \mbox{  otherwise, } \end{array}\right.\label{bmatrix}\end{equation}
give the transition probabilities for reaching  $E_{\x'}$,  $\x'= \x'_1,\x'_2,\dots \x'_\ka$ starting  from $E_\x$,  $\x= \x_1,\x_2,\dots \x_\ka$ after one step of classical evolution.
These matrices can be now ``quantized'' as follows. 
Let $\H\simeq\mathbbm{C}^p,$ be the  vector space of dimension $p$ with the scalar product $ \langle\cdot, \cdot\rangle$ and an
 orthonormal  basis $\{|j\rangle, j \in \{0\dots p-1\}\}$. Take $\u$ be a unitary transformation  on $\H$ such that  in the basis  above:
 \begin{equation} |\u_{i,j}|^2 =1/p \label{smallu},\qquad \u_{i,j}:=\langle i|\u|j\rangle.\end{equation}
  (One possible choice   for the matrix $\u_{i,j}$ is provided by the $p$-dimensional discrete Fourier transform.)  With each partition $\M_\ka$ we  now associate $N_\ka$-dimensional  Hilbert  space:
\[ \H_\ka= \underbrace{\H\otimes\H\otimes\dots\otimes\H}_\ka.\]
Using an orthonormal basis in  $\H_\ka$  given by the vectors:
\begin{equation*}
|\x\rangle:=|\x_1\rangle\otimes
 |\x_2\rangle\otimes\dots\otimes |\x_{\ka}\rangle, \qquad \x = \x_1\dots \x_{\ka}, \,\, \x_{i}  \in \{0\dots p-1\},
\end{equation*}
one  defines the unitary transformation $\bar{U}_\ka$ as:
\begin{equation} 
\bar{U}_\ka|\x\rangle=|\x_2\rangle\otimes
 |\x_3\rangle\otimes \dots\otimes|\x_{\ka}\rangle\otimes \u|\x_{1}\rangle.\label{umatrixhom}
\end{equation}
and the corresponding adjoint:
\begin{equation} 
\bar{U}_\ka^*|\x\rangle= \u^*|\x_\ka\rangle \otimes
 |\x_1\rangle\otimes |\x_2\rangle\otimes\dots\otimes|\x_{\ka-1}\rangle.\label{adumatrixhom}
\end{equation}
The action of $\bar{U}_\ka$ basically mimics the action of the  shift map. From this and property (\ref{smallu}) of $\u$ matrix it follows immediately that 
 $\bar{U}_\ka$  satisfies (\ref{umatrix}) and therefore, indeed, a quantization of $B_\ka$.
Note that if $\u$  is given by the discrete Fourier transform, the matrix  $\bar{U}_\ka$
  coincides with the evolution operator of the Walsh-quantized
 Baker map in \cite{AN1}. In that case $\bar{U}_\ka^2=-\mathbbm{1}$ and
  the spectrum  of $\bar{U}_\ka$ is highly degenerate. Note also that  $\u$ matrix in the definition (\ref{umatrixhom}) of $\bar{U}_\ka$ should not necessarily be a constant. 
More general
 construction is obtained if one takes $\u$ in the form 
\[\u(\x)= \exp(i\phi(\x))\u'(\x_2, \x_3\dots \x_{\ka}),\]
 where $\phi(\x)$ is a real
 function of $\x$ and  $\u'(\x_2, \x_3\dots \x_{\ka})$  is a unitary matrix depending on $\x_2, \x_3\dots \x_\ka$ and satisfying (\ref{smallu}).
\\

\noindent{\bf Maps with  non-uniform slopes.} Let us consider now the maps
 of the form 
\begin{equation}T_p(x)=p^{n_j} x \!\!\mod{1}, \qquad  \mbox{ for }
  x \in I_j, j=1,2\dots l, \label{nonuniformmap}\end{equation}
where  $n_j$ and  $p$ are integers such that $\sum_j^l p^{-n_j}=1$.  For a given $p$ we will use exactly the same representation $\x=\x_1\x_2\x_3\x_4\dots$,  $\x_i\in\{0,1\dots p-1\}$ for the point $x=0.\x$,  and the same set of the partitions $\M_\ka$ as for  the maps $\bar{T}_p$ with the uniform expansion rate. The action of  $T_p$ is again given by the shift map, but the size of the shift depends now on the point itself:
  \begin{equation}{T}_p:\x_1\x_2\x_3\x_4\dots\to \x_{n_j}\x_{n_j+1} \x_{n_j+2} \dots, \qquad  \mbox{ if }
 0.\x\in I_j, j=1,2\dots l.\end{equation}
The corresponding classical evolution
 matrix  for  the partition $\M_\ka$ is then  given  by
\begin{equation} B_\ka (\x,\x')=\left\{\begin{array}{lc} p^{-n_i} &\mbox{  if  } \dbrack{\x}\subseteq I_j \mbox{  and  } \x'_j=\x_{n_i+j}, \qquad j=1,\dots \ka-n_i\\
0 & \mbox{  otherwise.  } \end{array}\right.\label{bmatrixnonhom}\end{equation}
 It is not difficult  now  to ``quantize'' these  matrices using exactly the same Hilbert space as in the uniform case.   For each state $ |\x\rangle$, $\x=\x_{1}\dots \x_{\ka}$ such that  $\dbrack{\x}\subseteq I_j$,   define the action of   $U_\ka$ on $ |\x\rangle$ by:

\begin{equation} 
U_\ka |\x\rangle=|\x_{n_j+1}\rangle\otimes
  \dots \otimes|\x_{\ka}\rangle\otimes \u_{n_j}|\x_{n_j}\rangle \otimes \u_{n_j-1}|\x_{n_j-1}\rangle \otimes
 \dots  \otimes\u_1|\x_{1}\rangle, \label{umatrixnonhom}
\end{equation}
where all the matrices $\u_i$, $i=1,\dots n_j$   satisfy (\ref{smallu}).  It follows straightforwardly   from the  definition that $U_\ka$ is unitary and fulfills (\ref{umatrix}), thereby it is a ``quantization'' of $B_\ka$. As for the maps with uniform slopes, the matrices $\u_i$ do not need, in fact, be constant but could depend on $\x_{n_{\max}}, \x_{n_{\max}+1} \dots \x_\ka$, $n_{\max}=\max_{j} n_j$ as well. \\

\noindent{\bf Example:} As an example of the above quantization construction consider the map $T_2=T_{\{2,4,4\}}$ (see fig. 1b) which will be a principle  model for us in what follows. Explicitly, for $\x=\x_1\x_2 \x_3\dots$, $\x_i\in\{0,1\}$   the action of $T_2$ on $ x=0. \x$ is given by 
\begin{equation} T_2 (x)=\left\{\begin{array}{ll} 2 x \!\!\mod 1 &\mbox{  if  } 0\leq  x \leq 1/2\\
4 x \!\!\mod 1 & \mbox{  if  } 1/2\leq  x \leq1. \end{array}\right.\label{map2}\end{equation}
For the vector space $\H_\ka=\H\otimes\dots\otimes\H$ ($\ka$ times), $\H\simeq \mathbbm{C}^2$
the  corresponding quantum evolution acts on  $|\x\rangle\in \H_\ka$ as:
\begin{equation} U_\ka |\x\rangle =\left\{\begin{array}{ll}|\x_{2}\rangle\otimes|\x_{3}\rangle\otimes\dots\otimes
 |\x_{\ka}\rangle\otimes \u_{1}|\x_{1}\rangle  &\mbox{  if  } \x_1=0\\
|\x_{3}\rangle\otimes\dots\otimes
  \u_{2}|\x_{2}\rangle\otimes \u_{1}|\x_{1}\rangle  &\mbox{  if  } \x_1=1. \end{array}\right.\label{qmap2}\end{equation}

\section{Quantization of observables}

 We  recall now   the  procedure for the quantization of observables  introduced in \cite{BKS}.
Let $\M_\ka$ be the partition of the unite interval into $N_\ka$ intervals $\{ E_i=\left[(i-1){N^{-1}_\ka},i{N^{-1}_\ka}\right], i=1,\dots N_\ka\}$ and let $\H_\ka\simeq\mathbbm{C}^{N_\ka}$ denote the corresponding Hilbert space. For each function $f\in L^2(I)$  the corresponding  quantum observable $\Op(f)$ is  given by the matrix, whose elements are 
 \begin{equation}
\Op(f)_{i,j}{:=}\delta_{i,j}\frac{1}{N_\ka}\int_{E_i}f(x)\, dx, \qquad  i,j=1,\dots N_\ka. \label{quantization}
\end{equation}

 Set $I_c$ be the circle corresponding to $I=[0,1]$ where  the endpoints $0$ and $1$ are identified. It will be assumed that $I_c$ is equipped with the standart Euclidian metric coming from $\mathbbm{R}$. In particular the distance $d(x,y)$ between two points $x,y\in I_c$ is defined by  $d(x,y):=\min\{|x-y|,  |x-y-1|\}$.   In the present work we will often deal with  a class of observables $f\in\Lip(I_c)$ which are Lipschitz continues on $I_c$. Recall that the  space  $\Lip(I_c)$ is equipped with the Lipschitz norm:
\begin{equation}
\|f\|_{\Lip}=\sup_{x\in I}|f(x)|+\sup_{x\neq y\in I} \frac{|f(x)-f(y)|}{d(x,y)}
\end{equation}
and $f\in\Lip(I_c)$ iff $\|f\|_{\Lip}$ is finite.
The definition (\ref{quantization}) 
  is strongly motivated by  the existence of  the  correspondence between  classical and quantum evolutions of observables (Egorov property).  In the context of quantized one-dimensional maps the Egorov property was  proved in \cite[Thm.~3]{BKS}  for  Lipschitz continues  observables  undergoing one step evolution. 
The following theorem  is  a straightforward  extension of that result up to the  time  $ \nE:=\lfloor\log N_\ka/\log\Lambda_{\max}\rfloor$  which is a sort of Ehrenfest time for the model. (Here and after $ \lfloor y\rfloor$ denotes  the largest integer smaller then $y$.) 

\begin{theor} Let $U=U_\ka$ be a quantum evolution operator for a quantizable one-dimensional map $T$ (satisfying Conditions 1) and let  $f$ be a Lipschitz continuous function on $I_c$, then 
\begin{equation}\|U^{-n} \Op(f)U^{n}-\Op(f\circ T^n)\|\leq D(T) \|f\|_{\Lip}\frac{\Lambda^n_{\max}}{N_\ka}.\end{equation}\label{egorov}
where  $D(T)$ is a constant independent of $n$  and $N_\ka$. 
\end{theor}
\begin{prf}
For $n=1$ the  following bound was proved in \cite{BKS}:
 \begin{equation}\|U^{-1} \Op(f)U-\Op(f\circ T)\|\leq \|f\|_{\Lip}\frac{ D(T)}{N_\ka}.\end{equation}
From this one immediately  gets for $n$ iterations: 
\begin{eqnarray}\|U^{-n} \Op(f)U^{n}-\Op(f\circ T^n)\|&\leq& 
\sum_{i=1}^n \|U^{-i} \Op(f\circ T^{n-i})U^{i}-U^{1-i}\Op(f\circ T^{n-i+1})U^{i-1}\|\nonumber\\
&\leq& \sum_{i=1}^n \frac{ D(T)}{N_\ka}\|f\circ T^{i-1}\|_{\Lip}
\leq D(T)\|f\|_{\Lip}\frac{\Lambda^n_{\max}}{N_\ka},
\end{eqnarray}
where we used the fact that  $f\circ T^{i}\in \Lip(I_c)$ and $\|f\circ T^{i}\|_{\Lip}\leq \Lambda^i_{\max}\|f\|_{\Lip}$.
\end{prf}
A  direct consequence of Theorem \ref{egorov} is the following bound on the commutators which will be of use in what follows.

\begin{prop} Let $f\in \Lip(I_c)$, $g\in \Lip(I_c)$ then 

\begin{equation}\| [U^{-n}\Op(f) U^{n}, \Op(g) ]\|\leq 2D(T)\|g\|_{\Lip}\|f\|_{\Lip}\frac{\Lambda^n_{\max}}{N_\ka}
.\label{commutanteq}\end{equation}\label{commutantpr}
\end{prop}
\begin{prf}
Since $\Op(f\circ T^n)$ commutes with  $\Op(g)$ one has by  Theorem~\ref{egorov}:
\begin{eqnarray*}\| [U^{-n}\Op(f) U^{n}, \Op(g) ]\|&=& \| [U^{-n}\Op(f) U^{n}-\Op(f\circ T^n), \Op(g) ]\|\nonumber\\
&\leq& 2 D(T)\|f\|_{\Lip}\| \Op(g)\|\frac{\Lambda^n_{\max}}{N_\ka}.
\end{eqnarray*}
\end{prf}

It is worth to notice that for  a certain class of observables the Egorov property turns out to be exact. Let $x_1, x_2$ be two points on the lattice $\beta(\M_\ka)$ then with  an interval $X=[x_1, x_2]\subset I$ we can associate projection operator $P_{X}:=\Op(\chi_{\scriptscriptstyle{X}})$, where $\chi_{\scriptscriptstyle{X}}$ is the characteristic function on the set $X$. For such operators one has the following result.
\begin{prop}
Let $X\subset I$ be an  interval (or union of intervals) such that all the endpoints $\beta(X)$ and $ \beta(T^{-1} X)$ belong to  $\beta(\M_\ka)$, then
\begin{equation} U^{-1}P_{X}U= P_{ T^{-1} X}.\label{exactegoroveq}
\end{equation}\label{exactegorov}
 \end{prop}
\begin{prf}
Written in the matrix form the left   side of (\ref{exactegoroveq}) is given by
\begin{equation} (U^{*}P_{ X}U)_{l,m}=\sum_{\{j|E_j\subseteq X \}}(U_{j,l})^*U_{j,m}, \label{lokal} \end{equation}
where $E_j$ denotes  $j$'s element of the partition  $\M_\ka$. Observe that when $E_j\subseteq X$, the elements  $({U}_{j,l})^*\neq 0$, ($U_{j,m}\neq 0$) only if $T(E_l)\subseteq X$ (resp. $T(E_m)\subseteq X$). On the other hand, if the last condition holds, one can extend the summation in (\ref{lokal}) to all  values of $j$. By the unitarity of $U$ it gives the right side of  
(\ref{exactegoroveq}).
\end{prf}
 For the class of maps $T_p$ the proposition above implies the exact correspondence between classical and quantum evolutions of some projection operators up to the times of order $\nE$.
\begin{cor}
Let $T_p$, be a map of the form (\ref{nonuniformmap}). Denote $U$  a quantization of $T_p$ acting on the  vector space $\H_\ka$ of the dimension   $N_\ka=p^\ka$. For a cylinder $\dbrack{\x}$ of the length $|\x|=m$ the evolution of the corresponding projection operator  $P_{\dbrack{\x}}$ is given by 
\begin{equation} U^{-n}P_{\dbrack{\x}}U^n= P_{ T^{-n}\dbrack{\x}} \qquad \mbox{ \rm{for all} } n+m\leq\nE .
\end{equation}\label{cylinderegorov}
 \end{cor}
\begin{rem} {\rm
Note that  by approximating   continues observables with projection operators and using  Proposition~\ref{exactegorov} it is possible, in principle, to obtain   an alternative proof of Theorem~\ref{egorov}.}
\end{rem}


\section{Metric entropy of semiclassical measures}

Let $ U_\ka:\H_\ka\to\H_\ka$, $\ka=1,\cdots \infty$ be a sequence of   unitary quantizations of a quantizable  map $T$ satisfying Conditions 1.  For a given   sequence of the eigenstates: $ \psi_\ka\in\H_\ka$, $ U_\ka\psi_\ka=e^{i\theta_\ka} \psi_\ka$,  the corresponding measures  $\mu_\ka$,  $\ka=1,\cdots \infty$ are  defined by eq.~(\ref{rietz}) through the Riesz representation theorem. We will be concerned with the possible outcome for semiclassical T-invariant measures $\mu=\lim_{\ka\to\infty}\mu_\ka$. 
 Following the
 approach  of \cite{AN1, An, AN2} we will consider the  metric entropy $H_{\KS}(T,\mu)$ of  $\mu$.  Below we recall some basic
 properties of  classical entropies and connect them to a certain type of quantum
 entropies.

Let $\pi=\bigvee^s_{i=1} \I_i$  be a certain
 partition of  $I$ into $s$ intervals. Given a  measure $\mu$ on $I$  the entropy function of $\mu$  with respect to the
 partition $\pi$ is defined  by 
\[h_{\pi}(\mu):=-\sum_i\mu(\I_i)\log(\mu(\I_i)).\]
More generally, one can consider the pressure function:
 \[p_{\pi,v}(\mu):=-\sum_i\mu(\I_i)\log(v^2_i\mu(\I_i)),\]
where the weights $v=\{v_i:i= 1,\dots s\}$ are given by a set of real numbers fixed  for  a given partition. Obviously,  if all $v_i$  equal to one, then $p_{\pi,v}$ is just the entropy  defined above.
An important feature of  $h_{\pi}(\mu)$ its subadditivity property. If
 $\pi=\bigvee^s_{i=1} \I_i$ and $\tau=\bigvee^{s'}_{i=1} \J_i$ are two partitions, then for the
 partition  $\pi\vee\tau$ consisting of the elements $\I_i\cap \J_j$ and a measure $\mu$ one has:
\begin{equation}
h_{\pi\vee\tau}(\mu)\leq h_{\pi}(\mu)+h_{\tau}(\mu). \label{subadditivity1}
\end{equation}

 Now consider dynamically generated refinements of $\pi$. Define
 $\varepsilon=\varepsilon_0\varepsilon_1\dots \varepsilon_{n-1}$,  be a
 sequence of the elements $\varepsilon_{i}\in\{1,\dots s\}$ of the  length
 $|\varepsilon|=n$. For any $n\geq 1$  set partition  $\pi^{(n)}=
 \bigvee_{|\varepsilon|=n} \dbrack{\varepsilon}$ of $I$ be  collection of the sets:
\[\dbrack{\varepsilon}:=T^{-(n-1)}\I_{\varepsilon_{n-1}}\cap
 T^{-(n-2)}\I_{\varepsilon_{n-2}}\cap \dots \I_{\varepsilon_{0}}.\]
Each  cylinder $\dbrack{\varepsilon}$ has a simple meaning as the set of the
 points with the same ``$\varepsilon$-future''  up to  $n$ iteration. 
One is interested in the  entropies for $T$-invariant  measures $\mu$
 with respect to the partitions $\pi^{(n)}$:
\[ h_{n}(\mu):=h_{\pi^{(n)}}(\mu)=-\sum_{|\varepsilon|=n}\mu(\dbrack{\varepsilon})\log(\mu(\dbrack{\varepsilon})).\]
If $\mu$ is  $T$-invariant, it follows (see e.g., \cite{Ke}) 
 by the subadditivity (\ref{subadditivity1})   that:
 \begin{equation}
h_{n+m }(\mu)\leq h_{n}(\mu)+h_{m}(\mu).\label{subadditivity3}
\end{equation}
For the entropy function this implies the existence of the limit: 
 \begin{equation}
H_{\pi }(T,\mu)=\lim_{n\to\infty}\frac{1}{n} h_{n}(\mu).\label{almostentropy}
\end{equation}
The metric (Kolmogorov-Sinai) entropy is then defined as the supremum
 over all finite measurable initial partitions $\pi$:
\[ H_{\KS}(T,\mu) =\sup_{\pi}H_{\pi }(T,\mu).\]
 It worth to notice that the above  supremum   is actually reached automatically if one starts
 from the generating partition (e.g., $\bigvee^l_{i=1} I_i$).

In the quantum mechanical framework one needs to define a quantum
 observable   reproducing $h_{n}(\mu)$ (resp. $p_{n,v}(\mu)$) in the
 semiclassical limit. Note that a measure of each set $\I_i$ can be written as the 
 average $\mu(\I_i)=\int \chi_{\scriptscriptstyle{\I_i}}(x)\,d\mu$ over the  classical observable
  $\chi_{\scriptscriptstyle{\I_i}}(x)$ which is the characteristic function of the set $\I_i$.
The quantum observable corresponding to  $\chi_{\scriptscriptstyle{\I_i}}$ is then simply
 projection operator $P_{i}:=P_{\scriptscriptstyle{\I_i}}=\Op(\chi_{\scriptscriptstyle{\I_i}})$ on the set $\I_{i}$.
Now we need to ``quantize''  the   
 refined partitions $\bigvee_{|\varepsilon|=n} \dbrack{\varepsilon}$.  The most straightforward approach would be considering
 quantization of observables $\chi_{\scriptscriptstyle\dbrack{\varepsilon}}$. A different scheme
 was suggested in \cite{AN2}. Instead of  taking classically
 refined observables $\chi_{\scriptscriptstyle\dbrack{\varepsilon}}$  and then quantizing them, one
  considers     a natural quantum dynamical
 refinement of the initial quantum partition. 
We will say that a sequence of operators $\hat\pi=\{ \hat\pi_i, i=1\dots s\}$ defines {\it quantum partition} of $\H$  if 
 they  resolve  the unity operator:
\[\mathbbm{1}_{\H}=\sum_{i=1}^s\hat\pi^*_i \hat\pi_i.\]   For a quantum partition $\hat\pi$ the {\it entropy} (resp. {\it pressure}) of a state $\psi\in\H$ is given by
\[\hat{h}_{\hat\pi}(\psi):=-\sum_{i=1}^s\|\hat\pi_i\psi\|^2\log(\| \hat\pi_i\psi\|^2), \qquad  \hat{p}_{\hat\pi,v}(\psi):=-\sum_{i=1}^s\|\hat\pi_i\psi\|^2\log(\| \hat\pi_i\psi\|^2 v^2_i).\] 
Now with  each set 
$\dbrack{\varepsilon}$ of $\pi^{(n)}$  one  associates  the operator  defined
 by: 
\begin{equation}P_{\varepsilon}:=P_{\varepsilon_{n-1}}(n-1)\dots  P_{\varepsilon_1}(1)
 P_{\varepsilon_0}(0), \qquad P_{\varepsilon_{i}}(p)=
 U^{-p}P_{\varepsilon_{i}}U^{p}.\label{partitionoperator}\end{equation}
As follows
 immediately from the definition of $ P_{\varepsilon}$, the  sets of the operators $ \hat{\pi}^{(n)}=\{P_{\varepsilon}, \, |\varepsilon|=n\}$,  $ \hat{\pi}^{*(n)}=\{P^*_{\varepsilon}, \, |\varepsilon|=n\}$ define quantum partitions of $\mathbbm{1}_{\H_\ka}$. Note that  $P^*_{\varepsilon}$ and $ P_{\varepsilon}$ differ only by the order of the  components $ P_{\varepsilon_i}(i)$ and both $ \hat{\pi}^{*(n)}$,  $ \hat{\pi}^{*(n)}$ correspond  to the same classical partition $ {\pi}^{(n)}$. For an eigenfunction $\psi_\ka\in\H_\ka$ of the operator $U_\ka$ let $\hat{h}_{  \hat{\pi}^{(n)}}(\psi_\ka)$, $\hat{h}_{  \hat{\pi}^{*(n)}}(\psi_\ka)$ be the corresponding entropies. After introducing the weight functions:
\[\hat\mu_\ka(\dbrack{\varepsilon}):= \|P_{\varepsilon}\psi_\ka\|^2, \qquad \hat\mu^*_\ka(\dbrack{\varepsilon}):= \|P^*_{\varepsilon}\psi_\ka\|^2 \]
for  the elements $\dbrack{\varepsilon}$ of the corresponding classical partition $\pi^{(n)}$, the ``quantum'' entropies of $ \psi_\ka$ can be written with a slight abuse of notation (in principle, $\hat\mu_\ka$, $\hat\mu^*_\ka$ are   not  measures but merely positive weight functions defined only on the elements of the partitions) as the classical entropy function of $ \hat\mu_\ka$, $ \hat\mu^*_\ka$:
\begin{eqnarray}&\hat{h}_{\hat{\pi}^{(n)}}(\psi_\ka)= h_{n}(\hat\mu_\ka),& \qquad  h_{n}(\hat\mu_\ka)=-\sum_{|\varepsilon|=n}\hat\mu_\ka(\dbrack{\varepsilon})\log\hat\mu_\ka(\dbrack{\varepsilon}) \nonumber\\
&\hat{h}_{\hat{\pi}^{*(n)}}(\psi_\ka)= h_{n}(\hat\mu^*_\ka),& \qquad  h_{n}(\hat\mu^*_\ka)=-\sum_{|\varepsilon|=n}\hat\mu^*_\ka(\dbrack{\varepsilon})\log\hat\mu^*_\ka(\dbrack{\varepsilon}).
\label{qentropy}\end{eqnarray}
Note that the weight functions $\hat\mu_\ka$, $\hat\mu^*_\ka$ are closely related to the measure $\mu_\ka$ induced by the eigenstate $\psi_\ka$. For a finite $|\varepsilon|=n$ by the   Egorov property both  $\hat\mu_{\ka}(\dbrack{\varepsilon})$ and $\hat\mu^*_{\ka}(\dbrack{\varepsilon})$  equal to  $\mu_{\ka}(\dbrack{\varepsilon})$ up
 to   semiclassically small errors. Hence in the semiclassical limit: 
\begin{equation}
\lim_{\ka\to\infty}h_{n}(\hat\mu_\ka)=\lim_{\ka\to\infty}h_{n}(\hat\mu^*_\ka)=h_{n}(\mu),
\end{equation}
 where  $\mu= \lim_{\ka\to\infty}\mu_\ka$ is the corresponding semiclassical measure. To extract from $h_{n}(\mu)$ the metric entropy $H_\KS(T,\mu)$  of the measure $\mu$ it is necessary to apply the classical limit (\ref{almostentropy}). In  complete analogy,  the quantum pressures of  $\psi_\ka$: 
\begin{eqnarray}&\hat{p}_{\hat{\pi}^{(n)},v}(\psi_\ka)=p_{n,v}(\hat\mu_\ka),& \qquad  p_{n,v}(\hat\mu_\ka)=-\sum_{|\varepsilon|=n}\hat\mu_\ka(\dbrack{\varepsilon})\log\left(\hat\mu_\ka(\dbrack{\varepsilon}v^2_{\varepsilon}\right)\nonumber\\
&\hat{p}_{\hat{\pi}^{*(n)},v}(\psi_\ka)= p_{n,v}(\hat\mu^*_\ka),& \qquad  p_{n,v}(\hat\mu^*_\ka)=-\sum_{|\varepsilon|=n}\hat\mu^*_\ka(\dbrack{\varepsilon})\log\left(\hat\mu^*_\ka(\dbrack{\varepsilon}v^2_{\varepsilon}\right)
\end{eqnarray}
converge  in the  limit $\ka\to\infty$ to the classical pressure $p_{n,v}(\mu)$ of  $\mu$.

Note that so far we defined operators $P_i$ as quantizations of the characteristic functions of the intervals $\I_i$. Since $\chi_{\scriptscriptstyle \I_i}\not\in \Lip(I_c)$ one can not directly
 apply Theorem \ref{egorov} to the operators $P_{\varepsilon}$. For maps $T_p$ this can be circumvented by applying Conjecture \ref{cylinderegorov} instead. However, for general maps Proposition \ref{exactegorov} would imply the Egorov property only up to certain times usually shorter than $\nE$. In order to remedy this
 problem one can  consider a smoothened version   $\chi^{(\delta)}_{\scriptscriptstyle{\I_i}}(x)\in\Lip(I_c)$ of the 
 characteristic function. For the interval $\I_i=[\beta_-(\I_i), \beta_+(\I_i)]$  the function  $\chi^{(\delta)}_{\scriptscriptstyle{\I_i}}(x)$ is $1$ inside of the interval 
  $\I^{(\delta)}_i=[\beta_-(\I_i)+\delta, \beta_+(\I_i)-\delta]\subset\I_i$ and  smoothly  decaying to $0$ outside of $\I^{(\delta)}_i$ in a way that
\[1=\sum_{i=1}^s\left(\chi^{(\delta)}_{\scriptscriptstyle{\I_i}}\right)^2.\]
 The corresponding quantum  observables $P_{i}=\Op(\chi^{(\delta)}_{\scriptscriptstyle{\I_i}})$, $i=1,\dots s$ 
  then resolve the unity operator  and thereby the operators $P_{\varepsilon}, P^*_{\varepsilon}$, $|\varepsilon|=n$ defined by eq.~(\ref{partitionoperator}). Using quantum partition $\hat\pi^{(n)}_\delta=\{P_{\varepsilon}, |\varepsilon|=n\}$,  $\hat\pi^{*(n)}_\delta=\{P^*_{\varepsilon}, |\varepsilon|=n\}$ we  can define now by (\ref{qentropy}) the ``smoothened'' version $\hat h_{\hat\pi^{(n)}_\delta}(\psi_\ka)$,  $\hat h_{\hat\pi^{*(n)}_\delta}(\psi_\ka)$  of the quantum entropy (resp. pressure) of $\psi_\ka$.  After taking  the limits:
\[\lim_{\delta\to 0}\lim_{\ka\to\infty}\hat h_{\hat\pi^{(n)}_\delta}(\psi_\ka)=\lim_{\delta\to 0}\lim_{\ka\to\infty}\hat h_{\hat\pi^{*(n)}_\delta}(\psi_\ka)=h_n(\mu)\]
one reveals  (assuming that $\mu$ does not charge the boundary points  $\beta_\pm(\dbrack{\varepsilon})$ of the elements of the partition $\pi^{(n)}$) the entropy  of the semiclassical measure $\mu$.  In what follows, depending on the context,   we will use either  ``smooth''  ($\delta>0$) or ``sharp'' ($\delta=0$) versions of the quantum partitions $\hat\pi^{(n)}_\delta $, $\hat\pi^{*(n)}_\delta$. To simplify notation we will make use of the same symbol  $P_{\varepsilon}$ for the partition's elements in both cases but will state explicitly whether  it is of ``smooth``   or ``sharp``  type.  Also, for the sake of convenience we will fix throughout the paper the initial classical partition  to be $\pi=\pi^{(1)}=\bigvee_{i=1}^{l}I_i$.

\section{Bound on metric entropy}

The main purpose of this section is to  prove  the bound (\ref{firstresult}) on the
 possible values of $H_{\KS}(T,\mu)$. In what follows  we will closely follow
 the approach developed in \cite{AN2, AKN} for Anosov geodesic flows. The main
 technical tool is a variant of entropic uncertainty relation first
 proposed in \cite{De,Kr} and later generalized and proved in  \cite{MU}. Here we will make use of  a
 particular case of the statement  appearing in   \cite{AN2, AKN}. 

\begin{theor} {\rm (Entropic Uncertainty Principle \cite[Thm.~6.5]{AN2}.)} Let $\hat{\pi}=\{\hat{\pi}_i\}_{i=1}^s$, $\hat{\tau}=\{\hat{\tau}_i\}_{i=1}^{s'}$,   be two partitions of
 unity operator $\mathbbm{1}_\H$ on a complex Hilbert space $(\H,\langle.,.\rangle)$ and let $v=\{v_i\}_{i=1}^s$, $w=\{w_i\}_{i=1}^{s'}$ be the   families of the associated weights. For any normalized
 $\psi\in\H$ and any isometry $\U$ on $\H$ the corresponding pressures satisfy:
\begin{equation}
\hat{p}_{\hat{\pi},v}(\psi)+\hat{p}_{\hat{\tau},w}(\U\psi)\geq -2\log(\sup_{j,k}v_j w_k\|\hat{\pi}_j
 \U\,\hat{\tau}^*_k\|).\label{uncertaintyeq}
\end{equation}\label{uncertainty}
\end{theor}
In what follows we will use Theorem \ref{uncertainty} for the Hilbert space $\H_\ka$, quantum partitions $\hat{\pi}=\{P_{\varepsilon},  |\varepsilon|=n\}$, $\hat{\tau}=\{P^*_{\varepsilon},  |\varepsilon|=n\}$, defined by (\ref{partitionoperator}) as ''quantizations`` of the classical partition $\pi^{(n)}$, $\pi=\bigvee_{i=1}^{l}I_i$  and the corresponding  weights $v_{\varepsilon}=w_{\varepsilon}=\prod^{n-1}_{i=0}\Lambda^{-1/2}_{\varepsilon_i}$, $\varepsilon=\varepsilon_0\dots\varepsilon_{n-1}$. Furthermore, the isometry $\U$ will be the unitary transformation  $(U_{\ka})^{n}$ and the normalized state $\psi$ will be an  
  eigenstate $\psi_{\ka}$ of $U_{\ka}$.  With such a choice     the left side of (\ref{uncertaintyeq}) reads as:
\[p_{n,v}(\hat{\mu}_{\ka})+p_{n,v}(\hat{\mu}^*_{\ka})=
 -\sum_{|\varepsilon|=n}\hat{\mu}_{\ka}(\dbrack{\varepsilon})\log(\hat{\mu}_{\ka}( {\dbrack{\varepsilon}}) v^2_{\varepsilon})+
\hat{\mu}^*_{\ka}(\dbrack{\varepsilon})\log(\hat{\mu}^*_{\ka}( {\dbrack{\varepsilon}}) v^2_{\varepsilon}).
 \]
Thus, in order to bound
  $p_{n,v}(\mu_{\ka})$  from below   we need
 an estimation on the right hand side of (\ref{uncertaintyeq}). This amounts
 to the control over the elements:

 \[\|P_{\varepsilon}U^{n}P_{\varepsilon'}\|=\|\P_{\varepsilon\varepsilon'}\|, \qquad \P_{\varepsilon}=UP_{\varepsilon_0}UP_{\varepsilon_1}\dots UP_{\varepsilon_{n-1}}, \,\,\, \mbox{ where } \,\,\,  U=U_\ka.\]
The following proposition gives  the required estimation. 

\begin{prop} Let
 $\P_{\varepsilon}=UP_{\varepsilon_0}UP_{\varepsilon_1}\dots UP_{\varepsilon_{n-1}}$, then  \begin{equation}\|\P_{\varepsilon}\|\leq e^{nc\delta}
 N_\ka^{1/2}\prod_{i=1}^{n}\Lambda^{-1/2}_{\varepsilon_{i}}, \label{Nalinieq}\end{equation}
 where $c$ is a constant and $\delta$ is the smoothening parameter in the definition of   $P_{\varepsilon_{i}}$'s. \label{Nalini}
 \end{prop}

\begin{prf} 

For any  $v\in\H_\ka$, the absolute values of the components of the vector $v'=UP_{\varepsilon_{i}}v$ satisfy
the bound
\[ |v'_i|\leq(\Lambda^{-1/2}_{\varepsilon_{i}}+2\delta) \max_{i=1,\dots N_\ka}|v_i|.\]
Applying this inequality $n$ times one gets for the components of the vector  $v^{(n)}=\P_{\varepsilon}v$:
\[|v^{(n)}_i|\leq \left(\prod_{i=1}^{n}\Lambda_{\varepsilon_{i}}\right)^{-1/2}\left(1+2\Lambda^{1/2}_{\max}\delta\right)^n \max_{i=1,\dots N_\ka}|v_i|.\]
From this the desired estimation follows immediately with $c=2 \Lambda^{1/2}_{\max}$.
\end{prf}

 The entropic uncertainty principle  together with Proposition \ref{Nalini} then  give   the  bound on the pressure of $\psi_\ka$:   
\begin{equation} 
p_{n,v}(\hat{\mu}_{\ka})+p_{n,v}(\hat{\mu}^*_{\ka})\geq -2\log \left( e^{nc\delta}
 N_\ka^{1/2} \right),\label{quantumbound}
\end{equation}
which can be also written as 
\begin{equation}h_{n}(\hat{\mu}_{\ka})+h_{n}(\hat{\mu}^*_{\ka})\geq- \sum_{|\varepsilon|=n}\left(\hat{\mu}_{\ka}(\dbrack{\varepsilon}) + \hat{\mu}^*_{\ka}(\dbrack{\varepsilon}) \right)\log v_\varepsilon -2\log \left(e^{nc\delta}
 N_\ka^{1/2} \right).\label{equantumbound}\end{equation}
Note that  such a bound  becomes nontrivial only for times $n$ when  $v^{-1}_\varepsilon= \prod_{i=1}^{n}   \Lambda^{1/2}_{\varepsilon_i}$ is comparable with  $N_\ka^{1/2}$. In other words,   $n$ should be  of the same order as the Ehrenfest time $\nE$.  For  shorter times (\ref{equantumbound}) would only imply that $h_{n}(\hat{\mu}_{\ka})+h_{n}(\hat{\mu}^*_{\ka}) >C_0$, where $C_0<0$ (which is completely redundant as $h_n$ is a positive function).

It is now tempting to use the inequality (\ref{equantumbound}) for $n=\nE$ to get a bound on the metric entropy.  Recall, however, that in such a case the   relevant partition used to define $h_{\nE}$    is of the  quantum size  $N_\ka^{-1}$.
On the other hand, the correct  order of the semiclassical and classical  limits in the definition of $H_{\KS}(T,\mu)$  
 requires a bound on the entropy function  for partitions of a finite (classical) size, independent of $\ka$.  
Thus in order to extract useful information from  (\ref{quantumbound},\ref{equantumbound}) it is necessary to connect the pressure $p_{\nE,v}(\hat{\mu}_\ka)$ for  the quantum time  $\nE$ with the   pressure $p_{n,v}(\hat{\mu}_\ka)$ for an  arbitrary  classical time  $n$ (independent of $\ka$).  To this end it has been suggested in \cite{AN1} to make  use  of the    
 subadditivity of the metric entropy. More
 specifically,  for a classical  invariant
 measure $\mu$ the subadditivity of the entropy function implies: 
\begin{equation*}p_{n+m, v}(\mu)\leq p_{n,v}(\mu)+p_{m,v}(\mu), \qquad
 \Rightarrow 
\end{equation*}
 \begin{equation}
 p_{m,v}(\mu)\leq q p_{n,v}(\mu)+p_{r,v}(\mu), \qquad m=qn+ r. \label{classicalsubadditivity}\end{equation}
This cannot be applied straightforwardly, as   the  weights $\hat{\mu}^*_{\ka}, \hat{\mu}_{\ka}$, in general, are not  invariant under the action of $T$.  However,   by virtue of the Egorov property (Theorem \ref{egorov}) the measures   $\mu_\ka(\dbrack{\varepsilon})$ of sufficiently large cylinders $\dbrack{\varepsilon}$ are still approximately  invariant. As a result,  for $n\leq \nE$ the functions $p_{n,v}(\hat{\mu}_{\ka}),p_{n,v}(\hat{\mu}^*_{\ka})$  turn out to be subadditive  up to a  semiclassical  error. In such a situation  one can exploit    the inequality (\ref{quantumbound}) in conjunction with the approximate subadditivity of $p_{n,v}(\hat{\mu}_{\ka}),p_{n,v}(\hat{\mu}^*_{\ka})$ in order to prove the bound (\ref{firstresult}). 

\subsection{\bf $T_p$ maps.} To see precisely how the above scheme works out  it is instructive  first to treat the maps $T_p$ which were defined  in Section 4. Here it will be convenient to use sharp version of the partition ($\delta=0$) as we can utilize Corollary \ref{cylinderegorov} instead of Theorem \ref{egorov}. In comparison to general maps,  $T_p$-maps have an advantage, since by  Corollary \ref{cylinderegorov}  $\mu_\ka(\dbrack{\varepsilon})= \hat{\mu}^*_\ka(\dbrack{\varepsilon})=\hat{\mu}_\ka(\dbrack{\varepsilon})$ if $ |\varepsilon|=m\leq\nE$ and the measures  $\mu_\ka(\dbrack{\varepsilon})$  of the sets  $\dbrack{\varepsilon}$, remain  exactly invariant under $T^{-n}$: 
\begin{equation}\mu_\ka(\dbrack{\varepsilon})=\mu_\ka(T^{-n}\dbrack{\varepsilon}), \qquad \mbox{ for } n+m\leq \nE.\label{semmeasureinv}
\end{equation}
From this immediately follows the desired  connection between the pressures for
partitions of classical and quantum sizes.

\begin{prop} Let $T_p$, $\mu_\ka$ and $p_{n,v}(\mu_\ka)$ be  as defined above, then for    $\nE=qn+r$, $q,n,r \in\mathbbm{N}$, $0\leq r<n$:
\begin{equation}p_{\nE,v}(\mu_\ka) \leq  q p_{n,v}(\mu_\ka)+p_{r,v}(\mu_\ka).\end{equation} \label{quantumsubadditivity}
 \end{prop}
\begin{prf} Straightforwardly follows from the subadditivity of $h_{n}$ and (\ref{semmeasureinv}).
\end{prf}
Equipped with the above proposition we can prove now the bound (\ref{firstresult}) on the metric entropy for maps $T_p$.

\begin{theor} Let $U_{\ka}$, $\ka=0,\dots\infty$ be a sequence of unitary quantizations of  a map $T_p$, and let  $\{\psi_\ka\}$  be  a sequence of
 their eigenstates.  Then  the corresponding
 limiting invariant measure $\mu=\lim_{\ka\to\infty}\mu_\ka$ satisfies:  
\begin{equation}H_{\KS}(T_p,\mu)\geq
  \sum_{j}\mu(I_j)\log\Lambda_{j}-\frac{1}{2}\log\Lambda_{\max}.\label{tensorialbound}
\end{equation}
 \end{theor}
\begin{prf} From the bound (\ref{equantumbound}) and Proposition \ref{quantumsubadditivity} it follows that the pressure  for the partition of an arbitrary fixed  size $0<n<\nE$ satisfies the inequality:
\begin{equation}
\frac{p_{n,v}(\mu_\ka)}{n}\geq-\frac{1}{2}\log\Lambda_{\max}-\frac{p_{r,v}(\mu_\ka)}{\nE}- \frac{r}{n}\frac{p_{n,v}(\mu_\ka)}{\nE }. \label{local5}
\end{equation}
Because $r$, $p_{r,v}$ are bounded for a fixed  $n$,  the last three terms in the righthand side of (\ref{local5}) vanish when  $\ka\to\infty$ and one gets:
\begin{equation}
\frac{p_{n,v}(\mu)}{n}\geq-\frac{1}{2}\log\Lambda_{\max}. \label{local6}
\end{equation}
To complete the proof it remains to notice that
\[p_{n,v}(\mu)=h_n(\mu)-\sum_{|\varepsilon|=n}\mu(\dbrack{\varepsilon})\log\left(\prod_{i=1}^n\Lambda_{\varepsilon_i}\right),\]
and 
\[\lim_{n\to\infty}\frac{1}{n}\sum_{|\varepsilon|=n}\mu(\dbrack{\varepsilon})\log\left(\prod_{i=1}^n\Lambda_{\varepsilon_i}\right)=\sum_{j}\mu(I_j)\log\Lambda_{j}\]
by Birkhoff's ergodic theorem.
\end{prf}

\subsection{General maps. } To extend the bound (\ref{tensorialbound}) to all   maps satisfying Condition 1 one needs  an analog of   Proposition \ref{quantumsubadditivity}  for a general  $T$.  Note that in order to make  use of the Egorov property up to the Ehrenfest time $\nE$, we need for a general $T$  a smoothened version ($\delta>0$) of the  projection operators  $P_\varepsilon$ which we adopt in that section. 
As follows from the  lemma below,  by virtue of the Egorov property
the measure   $\mu_{\ka}$ is  invariant up
 to  a semiclassically small error  till the time $\nE$. 

\begin{lemm} Let $\dbrack{\varepsilon}$, $\varepsilon=\varepsilon_{0}
 \varepsilon_{2}\dots\varepsilon_{m-1}$ be  cylinder of the length $m=|\varepsilon|$.  Then
\begin{equation}\hat{\mu}_{\ka}(\dbrack{\varepsilon})=\hat{\mu}_{\ka}(T^{-n}\dbrack{\varepsilon})+R_{n,m},
 \qquad |R_{n,m}|\leq n C(g,m) \Lambda^{m+n}_{\max}/N_\ka,\end{equation}
where the constant $C(m)$ depends only on $m$. The same result holds for $\hat{\mu}^*_{\ka}$.  \label{measureinvariance}
 \end{lemm}

\begin{prf}
This lemma can be proven  using exactly the same chain of arguments as for a similar result in the case of Anosov geodesic flows in \cite[Prop.~4.1]{AN2}.   For the sake of completeness, we outline the proof for $m=1$. By the
 definition  $\hat{\mu}_{\ka}$-weight of the set $T^{-n}\dbrack{\varepsilon}=\cup_{|\varepsilon'|=n}\dbrack{\varepsilon'\varepsilon_0}$, $ \varepsilon':=\varepsilon'_{0}
 \varepsilon'_{2}\dots\varepsilon'_{n-1}$ is given by: 
\begin{eqnarray}\hat{\mu}_{\ka}(T^{-n}\dbrack{\varepsilon_0})&=&\sum_{|\varepsilon'|=n}\langle \psi
 P^*_{\varepsilon'\varepsilon_0} P_{\varepsilon'\varepsilon_0}\psi\rangle=\sum_{|\varepsilon'|=n}\langle \psi
 \P^*_{\varepsilon'} (P_{\varepsilon_0})^2\P_{\varepsilon'}\psi\rangle\nonumber\\
&=&\sum_{|\varepsilon'|=n}\langle \psi
 \P^*_{\varepsilon'_{0},
 \varepsilon'_{1}\dots\varepsilon'_{n-2}}  P_{\varepsilon'_{n-1}} (P_{\varepsilon_0}(1))^2P_{\varepsilon'_{n-1}} \P_{\varepsilon'_{0}
 \varepsilon'_{2}\dots\varepsilon'_{n-2}}\psi\rangle,
\end{eqnarray}
where $P_{\varepsilon_i}(m)=U^{-m}P_{\varepsilon_i}U^{m}$.
 Since  the commutator  $ [(P_{\varepsilon_0}(1))^2, P_{\varepsilon'_{n-1}}]$ is bounded by
 Proposition \ref{commutantpr} and $\sum_{\varepsilon'_{n-1}\in\{1,\dots s\}} P^2_{\varepsilon'_{n-1}}=\mathbbm{1}$, it is useful to change the order of $P_{\varepsilon_0}(1)$   and $P_{\varepsilon'_{n-1}}$. The result is:
\begin{eqnarray*}\hat{\mu}_{\ka}(T^{-n}\dbrack{\varepsilon_0}) &=&\sum_{|\varepsilon''|=n-1}\langle \psi
 \P^*_{\varepsilon''} (P_{\varepsilon_0}(1))^2\P_{\varepsilon''}\psi\rangle+R^{(1)}_{n,1},\nonumber\\
 R^{(1)}_{n,1}&\leq& \|[(P_{\varepsilon_0}(1))^2, P_{\varepsilon'_{n-1}}]\|\sum_{|\varepsilon''|=n-1}\langle \psi
 \P^*_{\varepsilon''} (P_{\varepsilon_0}(1))^2\P_{\varepsilon''}\psi\rangle\leq\|[(P_{\varepsilon_0}(1))^2, P_{\varepsilon'_{n-1}}]\|,
\end{eqnarray*}
where $ \varepsilon'':=\varepsilon'_{0}
 \varepsilon'_{2}\dots\varepsilon'_{n-2}$ and we used $\sum_{|\varepsilon''|=n-1}P^*_{\varepsilon''}P_{\varepsilon''}=\mathbbm{1}$.  Repeating this procedure $n$ times one gets:
\begin{equation}\hat{\mu}_{\ka}(T^{-n}\dbrack{\varepsilon_0})=\langle \psi_{\ka}
  (P_{\varepsilon_0}(n))^2\psi_{\ka}\rangle+R_{n,1}=\hat{\mu}_{\ka}(\dbrack{\varepsilon_0})+R_{n,1}\end{equation}
with the  reminder  $R_{n,1}$  bounded by:
\begin{equation}
|R_{n,1}|\leq n \max_{a,0<i\leq n-1}\|[(P_{\varepsilon_0})^2 (i), P_{a}]\|.
 \end{equation}
The lemma then follows from Proposition  \ref{commutantpr}. The cases of $\hat{\mu}^*_{\ka}$ and $m>1$ are 
  treated analogously. 
\end{prf}
 Thanks to the lemma above we can  show now that $p_{n,v}( \hat{\mu}_{\ka})$, $p_{n,v}( \hat{\mu}^*_{\ka})$ are 
 semiclassically subadditive functions. 

\begin{prop}
Let $\psi_\ka$ be a normalized  eigenstate of $U_{\ka}$ and let
 $\hat{\mu}_{\ka}(\dbrack{\varepsilon})=\|P_{\varepsilon}\psi_\ka\|^2$ be the corresponding weight function, then for any $1>\alpha\geq 0$ and times $n$
 such that, $n+m\leq  (1-\alpha)\log N_\ka/\log\Lambda_{\max}$:
\begin{equation}
p_{n+m,v}(\hat{\mu}_\ka)\leq p_{n,v}(\hat{\mu}_\ka)+ p_{m,v}(\hat{\mu}_\ka) +R'_{m}, \qquad |R'_{m}| <  C'(m)\log(N_\ka) N_\ka^{-\alpha},\label{approxsubadditivityf}
  \end{equation}
where the constant $C'(m)$ does not depend on $N_\ka$. The same result holds  for the weight function $\hat{\mu}^*_{\ka}(\dbrack{\varepsilon})=\|P^*_{\varepsilon}\psi_\ka\|^2$:
\begin{equation}
p_{n+m,v}(\hat{\mu}^*_\ka)\leq p_{n,v}(\hat{\mu}^*_\ka)+ p_{m,v}(\hat{\mu}^*_\ka) +R'^*_{m}, \qquad |R'^*_{m}| <  C''(m)\log(N_\ka) N_\ka^{-\alpha}.
  \end{equation}
 \label{approxsubadditivitypr}
\end{prop}
\begin{prf}
The subadditivity property (\ref{subadditivity1}) of  the entropy function implies: 
\begin{equation}p_{n+m,v}(\hat{\mu}_\ka)\leq p_{n,v}(\hat{\mu}_\ka)+ p_{m,v}(T^{n}_*\circ\hat{\mu}_\ka), \qquad \left(T^{n}_*\circ\hat{\mu}_\ka\right)(\dbrack{\varepsilon}):=\hat{\mu}_\ka(T^{-n}\dbrack{\varepsilon}).
 \end{equation}
Furthermore, since  $\hat{\mu}_\ka$ is invariant up to a semiclassical error  the second term could be written as 
\begin{equation}p_{m,v}(T^{-n}\circ\hat{\mu}_\ka)=-\sum_{|\varepsilon|=m}\hat{\mu}_{\ka}(T^{-n}\dbrack{\varepsilon})\log\left( \hat{\mu}_{\ka}(T^{-n}\dbrack{\varepsilon})\prod_{i=1}^m\Lambda(\varepsilon_i)\right)
=p_{m,v}(\hat{\mu}_\ka)+R'_{m},
 \end{equation}
where $R'_{m}$ can be easily estimated using Lemma \ref{measureinvariance} and continuity of the function
 $x\log x$:
\[ |R'_{m}|\leq C_1(m) |R_{n,m}|.\]
Here the constant  $C_1(m)$ depends only on $m$ and the proposition follows immediately from the bound on $|R_{n,m}|$. The case of  $p_{n,v}(\hat{\mu}^*_\ka)$ is treated analogously.
\end{prf}
\begin{prftheorem2}  Precisely as for the maps $T_p$, we can make use of Proposition \ref{approxsubadditivitypr}  and inequality (\ref{quantumbound}) to  get  the bound  on the pressure for  finite times. Let $\nE^{\alpha}:=\lfloor(1-\alpha)\log N_\ka/\log\Lambda_{\max}\rfloor$, with  ${\alpha}$ being as in  Proposition \ref{approxsubadditivitypr}.  Fixing a number $n$ and using the decomposition $ \nE^{\alpha}=qn+r$, with $n,q,r\in\mathbbm{N}$, $r\leq n$ one gets from (\ref{approxsubadditivityf}):
\begin{equation}
p_{\nE^{\alpha},v}(\hat{\mu}_\ka)\leq q p_{n,v}(\hat{\mu}_\ka)+p_{r,v}(\hat{\mu}_\ka)+q|R'_{n}|,
\end{equation}
and a similar inequality for the pressures of $\hat{\mu}^*_\ka$.
Now,  (\ref{quantumbound}) at the time  $n=\nE^{\alpha}$ and    the above subadditivity property provide us with the following bound:
\begin{eqnarray}
\frac{p_{n,v}(\hat{\mu}_\ka)+p_{n,v}(\hat{\mu}^*_\ka)}{n} 
 \geq &-&\frac{\log\Lambda_{\max}}{(1-\alpha)}-\frac{p_{r,v}(\hat{\mu}_\ka) 
+ p_{r,v}(\hat{\mu}^*_\ka) }{\nE^{\alpha}}
-\left(\frac{r}{n}\right)\frac{p_{n,v}(\hat{\mu}_\ka)
+p_{n,v}(\hat{\mu}^*_\ka) }{\nE^{\alpha}}\nonumber\\
&-&\frac{(|R'_{n}|+|R'^*_{n}|)(1-r/\nE^{\alpha})}{n} -2c\delta, \label{local7}
\end{eqnarray}
which  after taking the semiclassical limit   $\ka\to\infty$ reads as
\begin{equation}
\frac{p_{n,v(\Lambda)}(\mu)}{n}\geq-\frac{1}{2(1-\alpha)}\log\Lambda_{\max}- 2c\delta.
\end{equation}
 Finally, it remains to relate the pressure to the corresponding entropy function and take the limits $n\to\infty$, $\alpha\to 0$, $\delta\to 0$.
\end{prftheorem2}

\section{Proof of Anantharaman-Nonnenmacher conjecture for  $T_p$ maps}

As we have shown in  the previous section, the method of N. Anantharaman and S. 
 Nonnenmacher can be  employed   for the  proof of
 the bound (\ref{firstresult}). However, exactly  as for  Anosov geodesics flows, such an
 approach  does not allow  to prove a stronger result  (\ref{mainresult}). 
Very roughly, the reason for this can be explained in the following way. For a generic map the entropy function $h_n(\mu_\ka)$ is a ``non-homogeneous'' quantity which contains contributions from the cylinders $\dbrack{\varepsilon}$ with different "expansion rates" $\Lambda_\varepsilon$.  The domain of validity for subadditivity of the entropy function  is determined by an entry (cylinder) with the largest expansion rate and thus,   restricted to the times $n\leq\nE$. On the other hand, the bound (\ref{quantumbound}) becomes informative for times $n\geq\bar{n}$, where $\bar{n}=\nE\frac{\log\Lambda_{\max}}{2\, \overline{\log\Lambda}}$, $\overline{\log\Lambda}=\sum_{i=1}^l\mu(I_i)\log\Lambda_i$.
When the  expansion rate is highly  non-uniform  one is unable   to match 
 long ``quantum``  times $n>\bar{n}$ with  short ``classical`` times $n<\nE$, see fig.~3. This results in the bound  (\ref{firstresult}) which is clearly non-optimal (or even trivial in some cases). Below we  formulate a certain
 modification to the original strategy to overcome the problem.

\subsection{General idea}

 Speaking informally, the basic   idea here is to  ``homogenize'' 
   the original system, making  it uniformly expanding first and only then apply the method used in the
 previous section.
More specifically, we  consider the class of maps  $T= T_{p}$, defined in Section 3.2. In what follows we adopt the tower
 construction widely used in the theory of dynamical systems (see e.g.,
 \cite{Yo}). As we show  in the next subsection,  $T$ can
 be regarded as the first return map for  a certain  uniformly expanding dynamical system. Namely, the action of  $T$  on $I$ turns out to be equivalent to the action of the so-called  tower map
    $\widetilde{T}:\widetilde{I}\to\widetilde{I}$ 
  on  a subset  (``zero level'') of the  tower phase space $\widetilde{I}$. By  a
 standard construction for first return maps, any invariant measure
 $\mu$ for $T$ induces a  measure    $\widetilde{\mu}$ on $\widetilde{I}$
 invariant under   $\widetilde{T}$. The corresponding  metric entropies $
 H_{\KS}(\widetilde{T},  \widetilde{\mu})$, $ H_{\KS}(T,  \mu)$ are  then related to each other by Abramov's formula  and the entropic bound
  (\ref{mainresult}) turns out to be equivalent to:

\begin{equation}H_{\KS}(\widetilde{T},
  \widetilde{\mu})\geq\frac{1}{2}\log p.\label{boundgeneral}\end{equation}
Thus, in order to prove conjecture of   S. Nonnenmacher and N.
 Anantharaman for maps $T_{p}$ one needs to show (\ref{boundgeneral}) for the measure
 $\widetilde{\mu}$. 

It turns out that a pure classical construction above can be ``lifted'' to the quantum level. Recall  that  $\mu$ is a semiclassical measure  generated by
  eigenstates of a sequence $\{U_\ka\}$ of unitary quantizations   of
  $T$. A key observation is that 
   $\widetilde{\mu}$ is actually a semiclassical measure for a sequence $\{\widetilde{U}_\ka\}$ of quantizations of  $\widetilde{T}$.  In 
 Subsection~7.3 we show that for each sequence $\{\psi_\ka\}$ of the eigenstates of $\{U_\ka\}$
   generating in the semiclassical limit   the
 measure $\mu$ there exists  a sequence  $\{\Psi_\ka\}$ of  eigenstates of 
 $\{\widetilde{U}_\ka\}$   generating  the measure $\widetilde{\mu}$. This is
 schematically depicted by the following diagram: 
\begin{equation}
\xymatrix{
\mu_{\psi_\ka}\ar[d]^{\ka\to\infty} \ar@{=>}[r]^{Quantum} &  
 \mu_{\Psi_\ka}\ar[d]^{\ka\to\infty}\\
\mu=\mu\circ T^{-1} \,\,\,  \ar@{=>}[r]^{Classical} &
 \,\,\,\widetilde{\mu}=\widetilde{\mu}\circ\widetilde{T}^{-1} }\end{equation}
Since  $\widetilde{T}$ is a map with a uniform expansion rate one can
  apply the method used in the previous section in order to prove (\ref{boundgeneral}).
 From this the metric bound (\ref{mainresult}) follows immediately. 

\begin{rem} \rm{
As we would like to keep the exposition and notation below as simple as possible, we will first consider  in details  the  map  $T_{2}=T_{\{2,4,4\}}$ defined in (\ref{map2}). Most of the results  can then be straightforwardly extended to all other maps 
  $T_{p}=T_\Lbf$,
 $\Lbf=\{p^{n_1},\dots p^{n_l}\}$, where $p,n_i\in\mathbbm{N}$.
  }
\end{rem}

\begin{figure}[htb]
\begin{center}
\includegraphics[height=3.5cm]{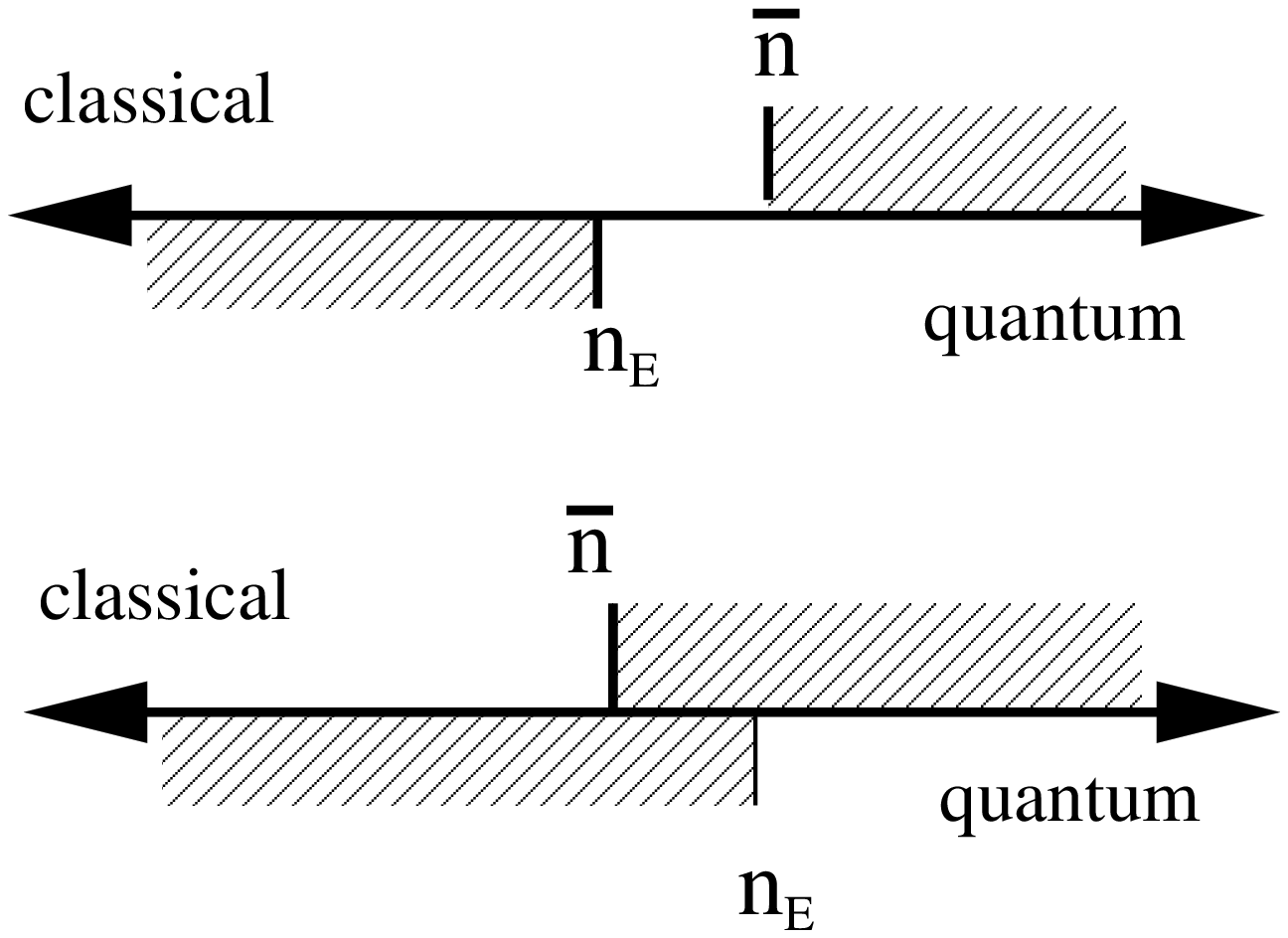} \hskip 1.5cm\includegraphics[height=3.5cm]{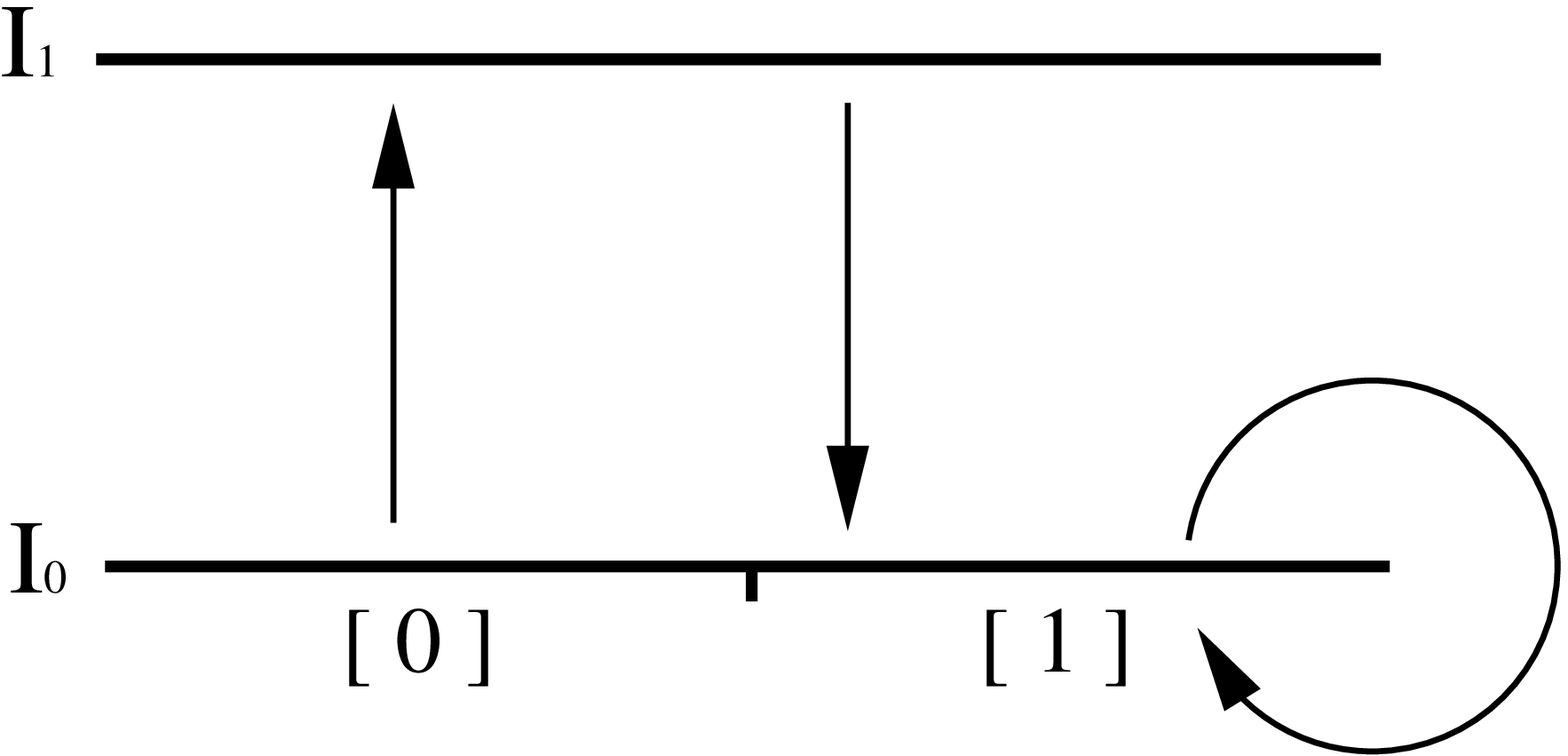} 
\end{center}
\caption{\small{ On the left down (up) is shown the case  when  (\ref{firstresult}) provides non-trivial (resp. trivial) bound on the metric entropy $ H_{\KS}(T,  \mu)$.   On the right is depicted  the tower for the map $T_{\{2,4,4\}}$.}} 
\end{figure}

\subsection{Classical  towers}

In what follows we  construct the tower dynamical system corresponding
 to the map $T:=T_{\{2,4,4\}}$ (as defined by eq. (\ref{map2})). To this end let us double the original 
 phase space and consider  the  set $\widetilde{I}:=I\times\{0,1\}$. We will referee to the sets
  $\widetilde{I}_0=\{(x,0), x\in I\}$, $\widetilde{I}_1=\{(x,1), x\in I\}$ as the first
 and second levels of the tower $\widetilde{I}=\Level_0\cup\Level_1$
 respectively. The tower map $\widetilde{T}: \widetilde{I}\to \widetilde{I}$
 is then defined by:

\begin{equation} \widetilde{T}(x,\eta)=\left\{ \begin{array}{rl}
  (\bar{T}(x), 0) & \mbox{ if } \eta=0, x\in[1/2,1] \mbox{ or } \eta=1 \mbox{ and any } x \\
(\bar{T}(x), 1) & \mbox{ if } \eta=0, x\in [0,1/2]\end{array}\right .\label{towermapp}
\end{equation}
where $\bar{T}:=T_{\{2,2\}}$ is the uniformly expanding  map corresponding to $T$.
Consider now  the first return map $\widetilde{T}_{\Level_0}$ on the set
 $\Level_0$.  It is then straightforward to see  that the action of
  $\widetilde{T}_{\Level_0}$ on $\Level_0\cong I$ coincides with the action
  of $T$ on $I$. In other words, $T$ can be regarded as the first
 return map for the lowest level of the tower (see fig. 3). 

Given an invariant measure $\mu$ for $T$ (equivalently for  
 $\widetilde{T}_{\Level_0}$) one can  construct (using a standard procedure, see
 e.g., \cite{Ke}, \cite{Lu})  the probability measure $\widetilde{\mu}$ which is
 invariant under the  tower map  $\widetilde{T}$. Precisely,  for a
 set $A\subseteq I$ one defines the measures of the sets $(A\times\{0\})$,
  $(A\times\{1\})$ by
\[\widetilde{\mu}(A\times\{0\})=\Gamma^{-1}\mu(A), \qquad
 \widetilde{\mu}(A\times\{1\})=\Gamma^{-1}\mu(\bar{T}^{-1}A\cap[1/2,1]),\] 
with the  normalization constant  $\Gamma=1+\mu([1/2,1])$. If $A=\dbrack{\x}$ is a cylinder set this can be rewritten as:
 \begin{equation}\widetilde{\mu}(\dbrack{\x}\times\{0\})=\Gamma^{-1}\mu(\dbrack{\x}), \qquad
 \widetilde{\mu}(\dbrack{\x}\times\{1\})=\Gamma^{-1}\mu(\dbrack{1\x})\label{towermeasureclass}.\end{equation}
 Since $\widetilde{\mu}$ is invariant under  $\widetilde{T}$ it makes
 sense to consider the corresponding metric entropy $H_{\KS}(\widetilde{T},
  \widetilde{\mu})$. An important observation is that
 $H_{\KS}(\widetilde{T},  \widetilde{\mu})$ is related to
 $H_{\KS}(T,  \mu)$. As $T$ is the  first return map
 for  $\Level_0$, and $\mu(\Level_0)=\Gamma^{-1}$, by Abramov's formula (see e.g., \cite{Ke}) one gets:
 \begin{equation} H_{\KS}(T,  \mu)=\Gamma
 H_{\KS}(\widetilde{T},  \widetilde{\mu}).\label{entropyconnection1}\end{equation}

Having an invariant measure $\widetilde{\mu}$ on $\widetilde{I}$ it is  possible in turn  to construct a  measure $\bar{\mu}$ on $I$ which is invariant under the homogeneous map $\bar{T}$. Let $\pii:\widetilde{I}\to I$ be a natural projection  on the tower: $\pii(x, \eta)=x$, for all $x\in I$, $\eta=\{0,1\}$. As 
\[\pii\circ \widetilde{T}=\bar{T}\circ \pii,\]
it follows immediately that the measure
\begin{equation}
\bar{\mu}:=\widetilde{\mu}\circ \pii^{-1}\label{barmeasureclass}
\end{equation}
is invariant under  $\bar{T}$. Furthermore, the metric entropy of $\bar{\mu}$ turns out to be equal to the metric entropy of $\widetilde{\mu}$:
\begin{equation} 
H_{\KS}(\bar{T},  \bar{\mu})=H_{\KS}(\widetilde{T},  \widetilde{\mu}). \label{homentropy1}
\end{equation}
This equality can be deduced,  from a version of the Abramov-Rokhlin relative entropy formula in \cite{DH}. For the sake of completeness we give a simple proof of  (\ref{homentropy1}) in  the appendix of the paper.

The above construction allows a straightforward extension to
 the case of an arbitrary map of the form $T_{p}=T_{\Lbf}$,  where $\Lambda_j=p^{n_j},\,\, j=1,\dots l$ and  $1<p\in\mathbbm{N}$.   The tower phase space here is  defined as $\htower:=\max_{j=1,\dots l}\{n_j\}$ copies of $I$:
\begin{equation}\widetilde{I}=I\times\{0,1,\dots ,\htower-1\}\cong \cup_{j=1}^{l-1}\widetilde{I_j},\end{equation}
where the  set $\widetilde{I}_j=I\times\{j\}$ stands for $j$'s level of the tower. The tower map $\widetilde{T}_{p}:\widetilde{I}\to\widetilde{I}$ is then defined with the help of the uniformly expanding map $\bar{T}_{p}$ given by eq.~(\ref{uniformmap}). For each level $\eta\in \{0,1,\dots ,\htower-1\}$  define the corresponding ``jumping'' set by
 \[   \falltower_\eta:=\cup_{\{j| n_j=\eta\}}I_j,\]
then the action of the  map $\widetilde{T}_{p}$ is given by:
\begin{equation} \widetilde{T}_{p}(x,\eta)=\left\{ \begin{array}{lr}
  (\bar{T}_{p}(x), 0) & \mbox{ if }  x\in \falltower_\eta  \\
(\bar{T}_{p}(x), \eta+1) & \mbox{ if }  x\not\in \falltower_\eta.\end{array}\right.
\end{equation}
Such a definition implies that with each iteration a  point in the tower phase space  climbs one step upstairs  up to  the moment when it   reaches at some level $\eta$ the set  $\falltower_\eta$. Then it  ``jumps'' downstare to zero level  and the  process is repeated.

 It is now straightforward to see that the map $T_{p}$ coincides with the first return map of $\widetilde{T}_{p}$ for zero level $\widetilde{I}_0$  of the tower. As a result, starting from an invariant measure $\mu$ for $T_{p}$ one can easily construct the invariant measure $\widetilde{\mu}$ for the tower map $\widetilde{T}_{p}$. For a set $A\times \eta\subseteq\widetilde{I}$, with $A\subseteq I$ and level $\eta\in\{0,\dots \htower-1\}$ the corresponding measure is given by 
\begin{equation}
\widetilde{\mu}(A\times \eta)= \Gamma ^{-1}\sum_{\{k|n_k\geq \eta\}}\mu(\bar{T}^{-\eta}_{p}(A)\cap I_k),\end{equation}
where $\Gamma=\sum_{j=1}^{l}n_j\mu(I_j)$ is the average return time to zero level of the tower. 
Precisely as for the map $T_{\{2,4,4\}}$, one can also construct the measure  $\bar{\mu}$ invariant under the action of  $\bar{T}_{p}$. The corresponding metric entropies are then related by:
\begin{equation} 
H_{\KS}(T_{p},  \mu)=\Gamma H_{\KS}(\bar{T}_{p},  \bar{\mu})=\Gamma H_{\KS}(\widetilde{T}_{p},  \widetilde{\mu}). \label{homentropy}
\end{equation}

\subsection{Quantum  towers}
We are going now to consider the quantum analog of the above tower construction. \\

\noindent{\bf Construction.}
Let $U=U_\ka$ be a tensorial quantization  of the map $T=T_{\{2,4,4\}}$, acting on the Hilbert space $\H=\H_\ka$ of the dimension  $2^\ka=\mathrm{dim}(\H_\ka)$. We will assume that $U$ is of the form  (\ref{qmap2}). In that case  $U$ allows an obvious decomposition:
\begin{equation}
 U=\bar{U}P_{\dbrack{0}}+\bar{U}_1\bar{U}P_{\dbrack{1}}, \end{equation}
where $\bar{U}$  stands for  a tensorial quantization of the uniformly expanding map  $\bar{T}=T_{\{2,2\}}$ acting on the Hilbert space $\H$ and  $
 \bar{U}_1=\sigma \bar{U}$ with  the unitary   $\sigma$ given by the exchange operation of the  last two  symbols in $|\x_1\rangle\otimes \dots \otimes|\x_{\ka-1}\rangle\otimes
 |\x_{\ka}\rangle\in\H$:
\[\sigma |\x_1\rangle\otimes \dots \otimes|\x_{\ka-1}\rangle\otimes
 |\x_{\ka}\rangle=|\x_1\rangle\otimes \dots \otimes|\x_{\ka}\rangle\otimes |\x_{\ka-1}\rangle.
 \]
In addition to  $P_{\dbrack{0}}$, $P_{\dbrack{1}}$ it will be also convenient to use  the projection operators: 
\begin{equation}
P'_{\dbrack{0}}=\bar{U}P_{\dbrack{0}}\bar{U}^*, \qquad P'_{\dbrack{1}}=\bar{U}P_{\dbrack{1}}\bar{U}^*.\label{proj}\end{equation}
Explicitly their action  on the basis states of $\H$ is given by:
\[ P'_{\dbrack{j}}|\x_1\rangle\otimes \dots\otimes |\x_{\ka-1}\rangle\otimes
 |\x_{\ka}\rangle=|\x_1\rangle\otimes \dots \otimes|\x_{\ka-1}\rangle\otimes\u \Pb_j\u^*|\x_{\ka}\rangle,
 \]
where $\Pb_j|i\rangle=\delta_{i,j}|i\rangle$,  $i,j\in\{0,1\}$. It worth to notice that $P'_{\dbrack{j}}$'s commute with $\bar{U}_1$:
\begin{equation}
\bar{U}_1P'_{\dbrack{1}}= P'_{\dbrack{1}}\bar{U}_1, \qquad \bar{U}_1P'_{\dbrack{0}}= P'_{\dbrack{0}}\bar{U}_1. \label{comut}
\end{equation}
 
We define  now the ``tower'' Hilbert space  $\widetilde{\H}=\widetilde{\H}_0\oplus
 \widetilde{\H}_1$,  $\mathrm{dim}(\widetilde{\H})=2^{\ka}+2^{\ka-1}$  with  
\begin{equation}\widetilde{\H}_0:=\H, \mbox{ and } \widetilde{\H}_1:=\bar{U}P_{\dbrack{1}}\H\equiv
  P'_{\dbrack{1}}\H,
\end{equation}
corresponding to zero and first levels of the tower.  The scalar
 product on $\widetilde{\H}$ is defined in a standard way using the scalar product at each level.
 Namely for  $\Phi=(\phi_0,\phi_1)\in\widetilde{\H}$,
 $\Phi'=(\phi'_0,\phi'_1)\in\widetilde{\H}$, with $\phi_0,\phi'_0\in\widetilde{\H}_0$ and
 $\phi_1,\phi'_1\in\widetilde{\H}_1$:
\[\bigbrack{\Phi,
 \Phi'}=\langle\phi_0,\phi'_0\rangle+\langle\phi_1,\phi'_1\rangle.\]
An orthonormal basis in $\widetilde{\H}$ can be easily constructed from an orthonormal basis in $\H$. A convenient choice is provided by the vectors:
\begin{align} \Ee_{(\x,0)}&:=(|\x_1\rangle\otimes \dots \otimes|\x_{\ka-1}\rangle\otimes
 |\x_{\ka}\rangle,\, 0),& \x&=\x_1\dots \x_{\ka-1}\x_{\ka};\nonumber\\
 \Ee_{(\x,1)}&:= (0, \,|\x_1\rangle\otimes \dots
 \otimes|\x_{\ka-1}\rangle\otimes |1'\rangle),&  \x&=\x_1\dots \x_{\ka-2}\x_{\ka-1},\label{towerbasis}
\end{align}
where $|0'\rangle:=\u|0\rangle$, $|1'\rangle:=\u|1\rangle$ and $\x_i$, $i=1,\dots \ka$ (resp. $i=1,\dots \ka-1$) run over all possible sequences of $\{0, 1\}$.

In what follows we will consider  one-parameter family of tower evolution operators $
 \widetilde{U}_{\theta}:\widetilde{\H}\to\widetilde{\H}$ defined in the following way. For any
 $\Phi=(\phi_0,\,\phi_1)\in\widetilde{\H}$, with $\phi_0\in\widetilde{\H}_0$ and
 $\phi_1\in\widetilde{\H}_1$:
\begin{equation}
 \widetilde{U}_{\theta}\Phi:=(\bar{U}_1 P'_{\dbrack{1}}\phi_1+\bar{U}P_{\dbrack{0}}\phi_0,\, e^{i\theta} \bar{U}P_{\dbrack{1}}\phi_0).\label{evtower}\end{equation}
Correspondingly,  the  adjoint operation $
 \widetilde{U}_{\theta}^*:\widetilde{\H}\to\widetilde{\H}$ is given by:
 \begin{equation}
 \widetilde{U}^*_{\theta}\Phi=(e^{-i\theta}P_{\dbrack{1}}\bar{U}^*\phi_1+P_{\dbrack{0}}\bar{U}^*\phi_0,\,  P'_{\dbrack{1}}\bar{U}_1^*\phi_0).\label{adevtower}\end{equation}

\noindent{\bf Main properties.} It is straightforward to see that
  $\widetilde{U}_{\theta} \, \Phi,\,  \widetilde{U}_{\theta}^*\, \Phi\in \widetilde{\H}$
 and $\widetilde{U}_{\theta}$ is a unitary operation on
 $\widetilde{\H}$:
\begin{prop} Let $\widetilde{U}_{\theta},\widetilde{U}^*_{\theta}$ be as above, then
\[\widetilde{U}_{\theta}\,\widetilde{U}^*_{\theta}=\widetilde{U}_{\theta}^*\,\widetilde{U}_{\theta}=\mathbbm{1}.\]
\end{prop}
\begin{prf}Straightforward calculation using eqs.~(\ref{proj}, \ref{comut}).  \end{prf}
 Below we demonstrate that the Egorov property holds for  
 $\widetilde{U}_{\theta}$.  Specifically the short time  evolution of projection
 operators is prescribed by the classical evolution of the corresponding tower
 map.  

\begin{prop} Let $\dbrack{\x}\subset I$ be a cylinder of the length $m=|\x|<\ka-1$, then: 
 \begin{equation}\widetilde{U}^*_{\theta}\,(P_{\dbrack{\x}}\oplus
 0)\,\widetilde{U}_{\theta}=(P_{\dbrack{0}}P_{\bar{T}^{-1}\dbrack{\x}}\oplus
 P'_{\dbrack{1}}P_{\bar{T}^{-1}\dbrack{\x}}),\label{egorovtower1}\end{equation}
\begin{equation}\widetilde{U}^*_{\theta}\,(0\oplus
 P'_{\dbrack{1}}P_{\dbrack{\x}})\,\widetilde{U}_{\theta}=(P_{\dbrack{1}}P_{\bar{T}^{-1}\dbrack{\x}}\oplus
 0).\label{egorovtower2}\end{equation}
\end{prop}
\begin{prf}
In the matrix  representation the left side of (\ref{egorovtower1}) reads as:
\begin{eqnarray}\left( \begin{array}{cc}
P_{\dbrack{0}}\bar{U}^* & e^{-i\theta}P_{\dbrack{1}}\bar{U}^*\\
P'_{\dbrack{1}}\bar{U}_1^* & 0 \end{array}\right) 
\left( \begin{array}{cc}  P_{\dbrack{\x}} & 0\\
 0 & 0 \end{array}\right) 
\left( \begin{array}{cc}
\bar{U}P_{\dbrack{0}} & \bar{U}_1 P'_{\dbrack{1}}\\
e^{i\theta} \bar{U}P_{\dbrack{1}} & 0 \end{array}\right). \label{matrix}
\end{eqnarray}
By eqs.~(\ref{proj}, \ref{comut}) the off diagonal terms of  the above product are zeros, and the diagonal part:
\[\left( \begin{array}{cc}
P_{\dbrack{0}}\bar{U}^* P_{\dbrack{\x}}\bar{U}P_{\dbrack{0}}& 0\\
0 & P'_{\dbrack{1}}\bar{U}_1^*P_{\dbrack{\x}}\bar{U}_1 P'_{\dbrack{1}}\end{array}\right)\]
coincides with  the right side of  (\ref{egorovtower1}) by Corollary \ref{cylinderegorov} and  obvious equality: $\sigma P_{\dbrack{\x}}\sigma =P_{\dbrack{\x}}$. Eq.~(\ref{egorovtower2}) is then proved analogously. 
\end{prf} 

\begin{cor} Let
 $\widetilde{P}_{\bar{T}^{-n}\dbrack{\x}}=(P_{\bar{T}^{-n}\dbrack{\x}}\oplus P'_{\dbrack{1}}P_{\bar{T}^{-n}\dbrack{\x}})$ be  the  projection operator on the
 subset $(\bar{T}^{-n}\dbrack{\x},0)\cup (0,\bar{T}^{-n}\dbrack{\x})$, $|\x|=m$ of the tower. Then for
 all $n+m<\ka-1$:
\begin{equation}
\widetilde{U}^*_{\theta}\,\widetilde{P}_{\bar{T}^{-n}\dbrack{\x}}\,\widetilde{U}_{\theta}=\widetilde{P}_{\bar{T}^{-n-1}\dbrack{\x}}.
\end{equation}
\end{cor}

\noindent{\bf Eigenfunctions and semiclassical measures.} Given an eigenfunction $\psi$
 of the original evolution operator $U$, $U\psi=e^{i\theta}\psi$  we can 
  construct  the eigenfunction of  the  tower evolution operator
  $\widetilde{U}_{\theta}$. Precisely, the state:
\begin{equation}\Psi=(\psi,  \bar{U}P_{\dbrack{1}}\psi)/\Gamma_\psi^{1/2}, \qquad
  \Gamma_\psi=1+\langle\psi,P_{\dbrack{1}}\psi\rangle,\label{eigenstates}
\end{equation}
is the normalized  eigenstate of the operator $\widetilde{U}_{\theta}$:
  $\widetilde{U}_{\theta}\Psi=e^{i\theta}\Psi$, $\bigbrack{\Psi,\Psi}=1$. 

For any sequence of eigenstates $\{\psi_\ka\}$ of quantizations $\{U_\ka\}$ of $T$
 one obtains applying   (\ref{eigenstates})  the  corresponding sequence of the eigenstates
  $\{\Psi_\ka\}$  of the quantizations 
 $\{\widetilde{U}_{\theta_\ka}\}$ of the tower map $\widetilde{T}$. As a result, a sequence of semiclassical measures $\mu_\ka$
 on $I$ induces the sequence of semiclassical measures $\widetilde{\mu}_\ka$ on
 $\widetilde{I}$.  
For a cylinder $\dbrack{\x}\subset I$ the   measures $\widetilde{\mu}_\ka$ of the tower  sets
 $\dbrack{\x}\times  \{0\}$, $\dbrack{\x}\times  \{1\}$ are defined as:
\[\widetilde{\mu}_\ka(\dbrack{\x}\times  \{0\})=\bigbrack{\Psi_\ka,\, P_{\dbrack{\x}}\oplus 0 \,
 \Psi_\ka}, \qquad \widetilde{\mu}_k(\dbrack{\x}\times  \{1\})=\bigbrack{\Psi_\ka,\, 0\oplus
 P_{\dbrack{\x}} \, \Psi_\ka}.\]
By eq.~(\ref{eigenstates}) these measures  are related  to    the  measure $\mu_{\ka}$   of the set $\dbrack{\x}$:
  \begin{equation}\widetilde{\mu}_\ka(\dbrack{\x}\times  \{0\})=\Gamma^{-1}_\ka{\mu}_\ka(\dbrack{\x}),
 \qquad \widetilde{\mu}_k(\dbrack{\x}\times
  \{1\})=\Gamma^{-1}_\ka{\mu}_\ka(\dbrack{1\x}),\label{towermeasure}\end{equation}
where we set $\Gamma_\ka=\Gamma_{\psi_\ka}$.
Note that after taking the  limit $\ka\to \infty$ in (\ref{towermeasure}) one obtains eqs.~(\ref{towermeasureclass}), where
 $\widetilde{\mu}=\lim_{\ka\to\infty}\widetilde{\mu}_\ka$ is precisely the measure   of
 the classical tower obtained from the semiclassical measure $\mu=\lim_{\ka\to\infty}{\mu}_\ka$ by the procedure from the previous section. Also,  defining  the
 measure $\bar{\mu}_{\ka}$ on $I$  by
  \begin{equation}
\bar{\mu}_{\ka}(\dbrack{\x}):=\bigbrack{\Psi_\ka,\widetilde{P}_{\dbrack{\x}}\Psi_\ka}=\Gamma^{-1}_\ka\left({\mu}_\ka(\dbrack{\x})+{\mu}_\ka(\dbrack{1\x})\right),\end{equation}
one reveals in the semiclassical limit the measure
 $\bar{\mu}=\lim_{\ka\to\infty}\bar{\mu}_\ka$ related to  $\widetilde{\mu}$ by eq.~(\ref{barmeasureclass}).

We leave it to the reader to check that the above construction can be extended to all maps $T_p$.

\subsection{Proof of Theorem~\ref{thmainresult}}
Let us now prove the bound (\ref{mainresult}) for the map $T_2$.
\begin{theor}
Let  $\{U_\ka\}_{\ka=1}^{\infty}$ be a sequence of  tensorial quantizations of $T=T_{\{2,4,4\}}$. For a sequence  $\{\psi_\ka\}_{\ka=1}^{\infty}$ of eigenstates  $U_\ka\psi_\ka=e^{i\theta_\ka}\psi_\ka$    let $ \mu=\lim_{\ka\to\infty}\mu_{\ka}$ be the corresponding  semiclassical measure, then:
\begin{equation} H_{\KS}(T, \mu)\geq\frac{\mu(\dbrack{0})+2\mu(\dbrack{1})}{2}\log 2.\label{boundtheor1} 
\end{equation}\label{maintheorex}
\end{theor}

\begin{prf} 
   To prove (\ref{boundtheor1}) it is possible, in principle,  to follow precisely  the scheme described in the beginning of the section  i.e., to prove the bound on $H_{\KS}(\widetilde{T}, \widetilde{\mu})$ for the corresponding semiclassical measure $\widetilde{\mu}$ on the tower and then deduce the bound (\ref{boundtheor1}) using Abramov's formula.  From the technical point of view, however,  it turns out to be    easier  to prove an equivalent bound for the metric entropy $H_{\KS}(\bar{T}, \bar{\mu})$ of the measure $\bar{\mu}$. 

Let $\{\Psi_\ka\}_{\ka=1}^{\infty}$ be the sequence of the tower eigenstates corresponding to the sequence of $\psi_\ka$'s, and let $\hat{h}_{n}(\Psi_\ka)\equiv h_{n}(\bar{\mu}_\ka)$ be the entropy function for the corresponding  measures $\bar{\mu}_\ka$:
\begin{equation}  
h_{n}(\bar{\mu}_\ka)=-\sum_{|\x|=n}\bar{\mu}_\ka(\llbracket \x\rrbracket)\log\bar{\mu}_\ka(\llbracket \x\rrbracket) =-\sum_{|\x|=n} \|\widetilde{P}_{\dbrack{\x}}\Psi_\ka \|^2\log(\|\widetilde{P}_{\dbrack{\x}}\Psi_\ka \|^2). \label{barbarentropy}
\end{equation}
Then the metric entropy $H_{\KS}(\bar{T},\bar{\mu})$ is obtained after first applying the semiclassical limit:
\begin{equation}  
h_{n}(\bar{\mu})=\lim_{\ka\to\infty} h_{n}(\bar{\mu}_\ka).
\end{equation}
and then the classical limit:
\begin{equation}  
H_{\KS}(\bar{T},\bar{\mu})=\lim_{n\to\infty}\frac{1}{n} h_{n}(\bar{\mu}).
\end{equation}
To prove the bound on  $H_{\KS}(\bar{T},\bar{\mu})$ we will make use of the same scheme as in \cite{AN1}. The first step is to get the  bound on   the entropy function, when $n$ is of of the same order as $\ka$. This is provided by the following proposition. 
\begin{prop}Let $h_{n}(\bar{\mu}_\ka)$ be as in (\ref{barbarentropy}) and set $n=\ka-1$, then 
\begin{equation}h_{\ka-1}(\bar{\mu}_\ka)\geq \left( \frac{\ka-1}{2}-1\right) \log 2 .\end{equation}
\label{towerentropyboundpr1}
\end{prop}
\begin{prf}
We will use the Uncertainty Entropic principle (Theorem~\ref{uncertainty}) for the partitions:  $\pi=\tau=\{\widetilde{P}_{\llbracket \y\rrbracket}, \,\,|\y|=\ka-1\}$, weights: $v_{\y}=w_{\y}\equiv 1$ and isometry operation $\U= (\widetilde{U}_{\theta_\ka})^{\ka-1}$. Since  $\Psi_\ka$ is an eigenstate of  $\widetilde{U}_{\theta_\ka}$ it follows immediately from (\ref{uncertaintyeq}):
\begin{equation} 
h_{\ka-1}(\Psi_\ka)\geq-\log(\sup_{|\y|=|\y'|=\ka-1}\|\widetilde{P}_{\llbracket \y\rrbracket} (\widetilde{U}_{\theta_\ka})^{\ka-1} 
\widetilde{P}_{\llbracket \y'\rrbracket}\|).\label{internalinequality}
\end{equation}
Thus one needs to estimate the norm of the matrix $\C(\y,\y')=\widetilde{P}_{\llbracket \y\rrbracket}( \widetilde{U}_{\theta_\ka})^{\ka-1} 
\widetilde{P}_{\llbracket \y'\rrbracket}$. To this end let us calculate the matrix elements of $\C(\y,\y')$:
\[\bigbrack{\Ee_{(\x,i)},\C (\y,\y')\Ee_{(\x',i')}},\]
in the basis of orthogonal states (\ref{towerbasis}) with the parameters: $i,i'\in\{0,1\}$,
 $|\x|=\ka-1$, ($|\x'|=\ka-1$)  if  $i=0$ (resp. $i'=0$)   and   $|\x|=\ka$, ($|\x'|=\ka$)  if  $i=1$ (resp. $i'=1$). The action of the projection operator on the basis states is given by
\begin{equation} 
 \widetilde {P}_{\llbracket \y\rrbracket} \Ee_{(\x,i)}=\Ee_{(\x,i)}\left(\prod_{m=1}^{\ka-1}\delta_{\x_m,\y_m}\right).
\end{equation}
Hence for each pair  of $\y, \y'$ there exist at most two values of $\x$ and two values of  $\x'$  such that the matrix elements  $\bigbrack{\Ee_{(\x,i)},\C (\y,\y')\Ee_{(\x',i')}}$ are not zeros. From that follows: 
\begin{equation}\|\C (\y,\y')\|\leq 2 \max_{(\x,i),(\x',i') } |\bigbrack{\Ee_{(\x,i)},\C (\y,\y')\Ee_{(\x',i')}}|=2 \max_{(\x,i),(\x',i') } |\bigbrack{\Ee_{(\x,i)},(\widetilde{U}_{\theta_\ka})^{\ka-1}\Ee_{(\x',i')}}|.\label{someinequality}\end{equation}
Therefore, it remains to estimate the elements of the  operator   $(\widetilde{U}_{\theta_\ka})^{\ka-1}$ in the basis of $\{\Ee_{(\x,i)}\}$. To this end, let us notice that the action of  $\widetilde{U}_{\theta_\ka}$ on $\{\Ee_{(\x,i)}\}$ up to times $\ka$ closely connected to the action of the corresponding tower map $\widetilde{T}$ on the sets $\dbrack{\x}\times\{i\}$ of $\widetilde{I}$. Specifically, let $\Ee=(\widetilde{U}_{\theta_\ka})^{\ka-1}\Ee_{(\x,i)}$. Then, as  follows from eq.~(\ref{evtower}), depending on  $\x$, $i$  the state $\Ee$ might take the values $(e,0)$ or $(0,e)$, where 
\begin{equation}e=e^{iQ\theta_{\ka}}|\x'_{\ka}\rangle\otimes  \u|\x_{i_1}\rangle\otimes \u|\x_{i_2}\rangle\otimes\dots \otimes\u|\x_{i_{\ka-1}}\rangle.
\end{equation}
Here   $\x_{i_1}, \x_{i_2}\dots \x_{i_{\ka-1}}$ is some permutation of the original sequence $\x_1, \x_2\dots \x_{\ka-1}$ and   $Q$ is an integer number.  Since  $|\langle \x_j, \u \x_i\rangle|=1/\sqrt{2}$ for any pair $\x_i, \x_j\in\{0,1\}$,  
\[  |\bigbrack{\Ee_{(\x',i')},(\widetilde{U}_{\theta_\ka})^{\ka-1}\Ee_{(\x,i)}}|=|\bigbrack{\Ee_{(\x',i')}, \Ee}|\leq 2^{-\left(\frac{\ka-1}{2}\right)}.
\]
Together with (\ref{internalinequality}) and (\ref{someinequality}) this gives the proof of the proposition.
\end{prf}

The second necessary step is to connect  values $h_{\ka-1}(\bar{\mu}_\ka)$ of the entropy  at  quantum times of order $\ka$ to its values $h_{n}(\bar{\mu}_\ka)$ at  short fixed classical times $n$. 

\begin{prop}Let $h_{n}(\bar{\mu}_\ka)$ be as in (\ref{barbarentropy}), and let $\ka-1=qn+r$, $r\leq n$ where $n<\ka-1$ is a fixed (classical) time and $q$, $r$ are  integers,   then 
\begin{equation} \frac{1}{n}h_{n}(\bar{\mu}_\ka)\geq \frac{1}{\ka-1}h_{\ka-1}(\bar{\mu}_\ka)-\frac{n\log 2}{\ka-1}.\label{subadditivity}\end{equation}\label{towerentropyboundpr2}
\end{prop}
\begin{prf} To prove (\ref{subadditivity})  one makes use of the fact that   the  measure $\bar{\mu}_\ka$ is invariant under the transformation $\bar{T}^j$ up to certain times $j$. From the definition  of  $\bar{\mu}_\ka$ and eq.~(\ref{cylinderegorov}) it   follows  that for any cylinder $\dbrack{\x}$ of a length $|\x|=m$:
 \begin{equation}\bar{\mu}_\ka(\dbrack{\x})=\bar{\mu}_\ka(\bar{T}^{-n}\dbrack{\x}), \mbox{ for } m+n\leq\ka-1.\label{invariance}\end{equation}
Let $n,q, r$ be  as in the  conditions of the proposition.
Then the subadditivity property (\ref{subadditivity3}) of the entropy function  implies
  \[h_{\ka-1}(\bar{\mu}_\ka)\leq -\sum_{j=0}^{q-1}\sum_{|\x|=n}\bar{\mu}_\ka(\bar{T}^{-jn}\dbrack{\x})\log\bar{\mu}_\ka(\bar{T}^{-jn}\dbrack{\x})-\sum_{|\x|=r}\bar{\mu}_\ka(\bar{T}^{-qn}\dbrack{\x})\log\bar{\mu}_\ka(\bar{T}^{-qn}\dbrack{\x}),\]
and by eq.~(\ref{invariance}) this reads as
 \begin{equation}h_{\ka-1}(\bar{\mu}_\ka)\leq qh_{n}(\bar{\mu}_\ka)+h_{r}(\bar{\mu}_\ka).\end{equation}
Since $|h_{r}(\bar{\mu}_\ka)|$ is bounded from above by $n\log 2$ one gets immediately  the inequality (\ref{subadditivity}).
\end{prf}

\noindent{\it  End of the proof of Theorem~\ref{maintheorex}:} The final step is  to combine Propositions~\ref{towerentropyboundpr1} and~\ref{towerentropyboundpr2}:
 \begin{equation} \frac{1}{n}h_{n}(\bar{\mu}_\ka)\geq \frac{\log2}{2}-\frac{(n+1)\log 2}{\ka-1}, \mbox{ for all } n<\ka.\label{almost}\end{equation}
Taking in (\ref{almost}) first limit $\ka\to\infty$ and then   $n\to\infty$ gives:
\[H_{\KS}(\bar{T},\bar{\mu})\geq \frac{\log 2}{2},\]
which by (\ref{homentropy1}, \ref{entropyconnection1}) implies the bound:
\begin{equation}H_{\KS}({T}, \mu)\geq \frac{\Gamma\log2}{2}.\end{equation} 
Since $\Gamma=\mu(\dbrack{0})+2\mu(\dbrack{1})$ this gives  the  bound (\ref{boundtheor1}).
\end{prf}
Theorem~\ref{maintheorex} can be straightforwardly generalized to other one-dimensional maps with slopes given by powers of the same integer.

\begin{prfs} All the ingredients of the above construction can be straightforwardly extended from the map $T_{\{2,4,4\}}$ to a general map  $T_{p}$. In particular, starting from an invariant semiclassical measure $\mu$ of  $T_{p}$ one can construct the invariant semiclassical measure $\widetilde{\mu}$ of the corresponding  tower  map $\widetilde{T}_{p}$  and the invariant semiclassical measure $\bar{\mu}$ of the corresponding  uniformly expanding map $\bar{T}_{\Lambda}$. Repeating then all the previous steps of the present section  one can show  the bound:
 \[H_{\KS}(\bar{T}_{p},\bar{\mu})\geq \frac{\log p}{2}.\]
Since  the  metric entropies  $H_{\KS}(\bar{T}_{p},\bar{\mu}), H_{\KS}(\widetilde{T}_{p},\widetilde{\mu}), H_{\KS}({T}_{p},{\mu})$ are connected to each other one immediately gets
  \begin{equation}H_{\KS}(T_{p},\mu)\geq \Gamma\frac{\log p}{2},\end{equation}
where $\Gamma$ is the measure $\mu$ of the tower. Finally, it remains to check that $\Gamma$ gives the correct prefactor. 
\end{prfs}


\section{Explicit sequences of ``non-ergodic" eigenstates}
Below we construct some explicit sequences of eigenstates for maps $\bar{T}_p$, $T_p$ quantized as in Section 3.2. 
Having such sequences we can calculate the induced semiclassical measures and test  the bound (\ref{mainresult}) for the corresponding metric entropies.

\subsection{Maps with uniform slopes} 
Let us first  consider the map $\bar{T}_p$  with the uniform slope  $p$ whose quantization is given  by eq. (\ref{umatrixhom}).
Note that if $\u$ is given by the discrete Fourier transform matrix,  the evolution operator $\bar{U}_\ka$ and the corresponding  eigenstates  are precisely the same as for Walsh-quantized baker's map treated in \cite{AN1}.   For   a general $\u$  the construction can be carried out in an  analogous way.  Let   $w\in\H$ be an eigenstate of $\u$, then 
 \begin{equation}\psi^{(w)}_\ka= \underbrace{w \otimes w \otimes\dots w}_\ka, \qquad \psi^{(w)}_\ka\in\H_\ka\label{tensorialstate}\end{equation}
is the eigenstate of $\bar{U}_\ka$. The semiclassical measure $\mu_w$ corresponding to the sequence $ \psi^{(w)}_\ka$, $\ka=1,\dots\infty$ and the  associated metric  entropy  of $\mu_w$  can  be
 then easily calculated.  Assuming that $w=\sum_{i=0}^{p-1} w_i|i\rangle$, where $\{|i\rangle, i=0,\dots p-1\}$ is an orthonormal basis in $\H$, the $\mu_w$-measure  of the cylinder set $\dbrack{\x}$, $\x=\x_1\dots \x_m$  is given by: 
\begin{equation} 
\mu_{w}(\dbrack{\x})=\lim_{\ka\to\infty}\langle\psi_{\ka}^{(w)}P_{\dbrack{\x}}
 \psi_{\ka}^{(w)}\rangle=\prod_{i=1}^m |w_{\x_i}|^2.
\end{equation}
As  this is the product measure, 
one gets for the metric entropy:
\begin{equation}
H_{\KS}(\bar{T}_p,\mu_w)=-\sum_{i=0}^{p-1} |w_i|^2 \log(|w_i|^2).
\end{equation}

A more general class of eigenstates can be constructed by taking a set of states $\w:=\{w^{(j)}\in\H, j=0,\dots d-1\}$  cyclically related to each other: $\u w^{(j)}=w^{(j+1 \mod d)}$. Now 
define $\w_0:=w^{(0)}\otimes w^{(1)}\dots \otimes w^{(d-1)}$
and let  $\w_1, \w_2,  \dots \w_{d-1} $ be the vectors obtained from  $\w_0$ by cyclic permutation of its components, e.g., 
\[\w_i:=w^{(i\mod d)}\otimes w^{(1+i\mod d)}\dots \otimes w^{(d-1+i\!\mod d)},\qquad i=0,\dots d-1.\]
For each $\ka$  satisfying $\ka\!\!\mod d =0$ one  looks for   eigenstates of $\bar{U}_\ka$ in the form
\begin{equation}
 \psi_{\ka}^{(\w)}=\sum_{i=0}^{d-1} \C^{(\ka)}_i\,\underbrace{\w_i \otimes \w_i \otimes\dots \w_i}_{\ka/d}, \qquad \psi_{\ka}^{(\w)}\in\H_{\ka}.\label{multytensorialstate}
\end{equation}
The normalization condition $\|\psi_{\ka}^{(\w)}\|=1$ implies:
\[\sum_{i=0}^{d-1}\left(\C_i\right)^2=1, \qquad \C_i=\lim_{\ka\to\infty}|\C^{(\ka)}_i|.\]
When all  $\C^{(\ka)}_i$  are equal, one gets by (\ref{multytensorialstate}) the eigenstate of $\bar{U}_\ka$  . (Note that the eigenstates  (\ref{tensorialstate}) could be seen as a particular case of (\ref{multytensorialstate}) when $d=1$.)  The corresponding semiclassical measure is then given by the sum of the product measures
\begin{eqnarray} 
\mu_{\w}(\dbrack{\x})=\lim_{\ka\to\infty}\langle\psi_{\ka}^{(\w)}P_{\dbrack{\x}}\psi_{\ka}^{(\w)}\rangle=\sum_{i=0}^{d-1} (\C_i)^2 \mu^{(i)}_{\w}(\dbrack{\x}), \nonumber\\
 \mu^{(i)}_{\w}(\dbrack{\x})=\prod_{j=1}^m |w^{(i+j-1\!\!\!\mod d)}_{\x_j}|^2,\label{multytensorialmeasure}
\end{eqnarray}
where $ w^{(j)}_{i}$ is $i$'s component of the vector $ w^{(j)}$ in the basis $\{|i\rangle, i=0,\dots p-1\}$.
Although $\mu_{\w}$  is not a simple product measure, it is still possible to calculate the metric entropy explicitly:
\begin{equation}
H_{\KS}(\bar{T}_p,\mu_\w)=-\sum_{i=0}^{d-1}\sum_{j=0}^{p-1} |w^{(i)}_{j}|^2 \log(|w^{(i)}_{j}|^2).\label{homogeneousentropy}
\end{equation}
From  a simple application of Uncertainty Entropic Principle it follows that  $H_{\KS}(\bar{T}_p,\mu_w)\geq\frac{1}{2} \log p$ which is precisely the bound (\ref{firstresult})  (equivalent to (\ref{mainresult})   in that case). Furthermore,  for $\u$ given by the discrete Fourier transform and $d=1$ there exist vectors $w$  such that measures    $\mu_w$ saturate the above bound \cite{AN1}. 

Note that if all $w^{(i)}_{j}\neq 0$ the measures above are  supported on the whole $I$. As has been shown in  \cite{AN1} in the case when  $\u$ is the discrete Fourier transform matrices,   it is also possible to construct an entirely different class of exceptional sequences of eigenstates where  parts of the corresponding semiclassical measures are  localized on the periodic trajectories. This is due to the fact that  when $\u^n=\mathbbm{1}$ for some small integer $n$, the spectrum of $\bar{U}_\ka$ becomes highly degenerate. Since  no such degeneracies are expected  for quantized maps with non-uniform slopes, it seems that this type of  semiclassical measures can be constructed  only for the maps $\bar{T}_p$.  We refer the reader to  \cite{AN1}, \cite{DeFN}  for the details of the construction.

\subsection{Maps with non-uniform slope}

For maps $T_p$ we will look for sequences of  eigenstates having exactly  the same form  (\ref{multytensorialstate}, \ref{multytensorialstate}) as for the uniform case.  As we show, one can construct such  sequences by  choosing the matrices $\u_i$ and the constants $\C^{(\ka)}_i$ in an appropriate way.  Below we give several concrete examples of such a construction for the map (\ref{map2}) whose quantization is given by (\ref{qmap2}). \\

\noindent{\bf Example 1.} Let  $\u_1=\u_2=\u$ be two by two matrix satisfying $\u^2=-\mathbbm{1}$, $|\u(i,j)|=1/\sqrt{2}$, e.g., discrete  Fourier transform. 
Let $\u|1\rangle=:|e_+\rangle$. Since    $\u|e_+\rangle=-|1\rangle$ it can
 be easily seen that  for even $\ka$ 
 \begin{equation} \psi_{\ka}^{(1)}=
 \underbrace{|1\rangle\otimes|e_+\rangle\dots|1\rangle\otimes|e_+\rangle}_{\ka}
\end{equation}
is  an eigenstate of $U_{\ka}$.   For the sequence of states
 $\psi_{\ka}^{(1)}$ the induced semiclassical measure $\mu_\ka^{(1)}$ has entire support at the Cantor set. The
 metric entropy for this measure
can be easily calculated: $H_{\KS}(T_2, \mu^{(1)}_\ka)=\log 2$. Note that
  $H_{\KS}(T_2, \mu^{(1)}_\ka)$  saturates the bound (\ref{mainresult}) which in that case coincides with  (\ref{firstresult}).     \\

\noindent{\bf Example 2.} For the same map $T_2$  consider a
 slightly  different  quantization.  Let $\u$ be an arbitrary unitary
 matrix whose elements have modules $1/\sqrt{2}$ and let $w$  be one
 of its eigenvectors with the eigenvalue $e^{i\gamma}$. We now fix
 $\u_1$, $\u_2$ by the  conditions $\u_2=e^{-i\gamma}\u$, $\u_1=\u$. The state
 
\[ \psi_{\ka}^{(2)}= \underbrace{w \otimes w \otimes\dots
 w}_{\ka},\]
is then the  eigenstate of $U_\ka$. Denote  $\mu^{(2)}_{w}$ the corresponding semiclassical measure.
 Unlike the previous example, in general,  $\mu^{(2)}_{w}$ is  supported over
 all $I$.
For a given state $w=w_0|0\rangle+w_1|1\rangle$ the measures of the sets
 $\dbrack{\varepsilon_0}, \varepsilon_0=\{1,2,3\}$ are:
\[\mu^{(2)}_{w}(\dbrack{\varepsilon_0})=\lim_{\ka\to\infty}\langle\psi_{\ka}^{(2)}P_{\varepsilon_0}
 \psi_{\ka}^{(2)}\rangle=\left\{ \begin{array}{ll}p& \mbox{ for }
 \varepsilon_0=1\\
p q& \mbox{ for } \varepsilon_0=2\\
q^2& \mbox{ for } \varepsilon_0=3,\\
\end{array}\right.\]
where $|w_0|^2=p$, $|w_1|^2=q$. Since $\mu^{(2)}_{w}$ is a product measure 
  the corresponding metric entropy is given by: 
\[
 H_{\KS}(T_2, \mu^{(2)}_{w})=-\sum_{\varepsilon_0=\{1,2,3\}}\mu^{(2)}_{w}(\dbrack{\varepsilon_0})\log\mu^{(2)}_{w}(\dbrack{\varepsilon_0})=-(p\log p +pq\log (pq)+q^2\log q^2).\] 
Recall that $w$ is an eigenvector of a unitary  matrix  whose entries have the same modules. This
 restricts the possible values of $q$, $p=1-q$ to the interval
 $[(2-\sqrt{2})/4,(2+\sqrt{2})/4]$. As can be easily  checked for all values of $q,p$ in
  this interval the strict inequality (\ref{mainresult}) holds.   It worth to notice that this example allows straightforward generalization to all maps $T_p$.\\

\noindent{\bf Example 3.} It is also
 possible to construct eigenstates of $U_\ka$ using two state products: 
\begin{eqnarray*} \psi_{\ka}^{(3)}&=&\C^{(\ka)}_1\underbrace{ w^{(1)} \otimes
 w^{(2)} \otimes w^{(1)} \otimes w^{(2)}\dots w^{(1)} \otimes w^{(2)}}_{\ka}\\
 &+& \C^{(\ka)}_2\underbrace{ w^{(2)}\otimes w^{(1)}\otimes w^{(2)} \otimes w^{(1)}\dots  w^{(2)}
 \otimes w^{(1)}}_{\ka}.\end{eqnarray*} 
 Take
 \[\u_2=\u_1=\u,\qquad \u=\frac{1}{\sqrt{2}}\left( \begin{array}{cc}
 1 & e^{i\alpha} \\
 e^{-i\alpha} & -1
\end{array}\right), \qquad (\u)^2=\mathbbm{1}. \]
and set $\C^{(\ka)}_1=z\C^{(\ka)}_2$, $c=1+|z\sqrt{2}-1|^2$,
\[w^{(1)}=c^{-1/2}\left(|0\rangle + e^{-i\alpha}(z\sqrt{2}-1)|1\rangle\right), \qquad  w^{(2)}=c^{-1/2}\left(
 z|0\rangle +e^{-i\alpha} (\sqrt{2}-z)|1\rangle\right).\]
 It is easy to check that $\psi_{\ka}^{(3)}$ is the eigenstate of $U_\ka$ for any $z\in\mathbbm{C}$.   The resulting semiclassical measure $\mu_{z}^{(3)}$  is  the sum of two product measures (defined by eq.~(\ref{multytensorialmeasure})). Note that $\mu_{z}^{(3)}$ is symmetric under the inversion
 $z\to z^{-1}$.  Denote $p_{1,2}=|w^{(1,2)}_{2}|^2$, $q_{1,2}=|w^{(1,2)}_{1}|^2$.  As will be  shown  in the rest of the section,
  the metric entropy of $\mu_{z}^{(3)}$   can be explicitly calculated and it is given by
\[H_{\KS}({T_2, \mu_{z}^{(3)}})= -\frac{\Gamma}{2}\sum_{k=1,2}p_k\log
 p_k +q_k\log q_k,\]
where 
\[ \Gamma=2(\mu([10])+
 \mu([11]))+\mu([0])=1+\C_1 p_1+\C_2p_2
 .\] The plot in fig.~4 shows both the metric entropy and the bound (\ref{mainresult}): $\frac{\Gamma}{2} \log 2$ as functions of the real part of $ z$ for $\mathrm{Im}(z)=0$. 

{\it Some special cases: } 1) $|z|=1$. In this case $p_1=p_2$, $q_1=q_2$ and the
 resulting measures of the simple product type. Furthermore, both
 $w^{(1)}$ and $w^{(2)}$ are the eigenvectors of the same unitary matrix whose elements
 have equal modulus. Thus one actually, gets the measures of the same type
  as for one vector product states    $\psi_{\ka}^{(2)}$  in
 the previous example. 2) $z=0, z=\infty$. In that case either $\C_2$ or
 $\C_1$ vanishes and we get the states considered in  Example 1. 3)
 $z=\sqrt{2}, z^{-1}=\sqrt{2}$. In such a case $p_1=q_1=1/2$, $p_2=0,
 q_2=1$ and the metric entropy $H_{\KS}({T_2, \mu_{\scriptscriptstyle{\sqrt{2}}}^{(3)}})=\frac{2}{3}\log 2$ saturates the
 bound.    \\ 

\begin{figure}[htb]
\begin{center}
\includegraphics[height=5cm]{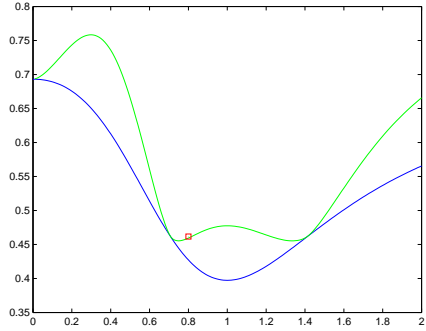}
\end{center}
\caption{\small{Metric entropy (green)  and the corresponding bound (blue) (\ref{mainresult}) for the semiclassical measure in Example 3 as function of $ \mathrm{Re} (z)$ when $  \mathrm{Im}(z)=0$. }} 
\end{figure}

The above examples can be generalized to other maps   $T_p$ to construct   d-state product  eigenstates of the type (\ref{multytensorialstate}).  More specifically, assume that  by an appropriate choice of constants $\C^{(\ka)}_i$  in  one can construct 
an eigenstate  $\psi_{\ka}^{(\w)}$ of the quantum evolution operator  $U_\ka$ (\ref{umatrixnonhom}) with $\u_i=\u$, for all $i$. It is instructive to see how the metric entropy of the corresponding semiclassical measures $\mu_{\w}$ can be calculated in general case.

 Note that   $\psi_{\ka}^{(\w)}$ being an eigenstate  of $U_\ka$, is in addition,  an eigenstate for  the operator $(\bar{U}_\ka)^d$, where $\bar{U}_\ka$ is the quantization  (\ref{umatrixhom}) of the map $\bar{T}_{p}$ with the uniform slope $p$.  Since  $(\bar{U}_\ka)^d$ is  also a quantization   of the map $\bar{T}_{p^d}$, the  semiclassical measure $\mu_{\w}$ turns out to be 
 invariant  both for $T_p$  and   $\bar{T}_{p^d}$ maps. The corresponding metric entropies $H_{\KS}(T_p,\mu_{\w}) $ and  $H_{\KS}(\bar{T}_{p^d},\mu_{\w}) $ can be connected to each other in the following way.  
Using   either $T_p$  or  $\bar{T}_{p}$ and the corresponding  dynamically generated    partitions,
one can encode any point $\zeta\in I$  according to its dynamical ``history`` in a two-fold way. The ''history`` 
with respect to $T_p$  and  $\bar{T}_{p}$ are   given  by the  sequences $\varepsilon(\zeta)=\varepsilon_0\varepsilon_{1 }\dots$,  $\varepsilon_i\in\{1,\dots l\}$ and  $\x(\zeta)=\x_1\x_2\dots $,  $\x_i\in\{0,\dots p-1\}$ respectively. Furthermore, each of these sequences   generates the set of cylinders: $\{\dbrack{\varepsilon_0\dots\varepsilon_{n }}, n=0,1,\dots\}$, $\{\dbrack{\x_1\dots \x_{n }}, n=1,2,\dots\}$  corresponding to the ''partial histories`` of the point evolution with regards to $T_p$  and  $\bar{T}_{p}$ respectively. 
The  Shannon-McMillan-Breiman theorem asserts then that for almost every  (with respect to $\mu_{\w}$) $\zeta\in I$ the metric entropy of  $T_p$  is given  by:
\begin{equation}
 H_{\KS}(T_p,\mu_{\w})=-\lim_{n\to\infty}\frac{1}{n}\log\mu_{\w}(\dbrack{\varepsilon_0\dots\varepsilon_{n}}).
\end{equation}
Analogously, using the second representation for the same point $\zeta$ one gets:
  \begin{equation}
 \frac{1}{d}H_{\KS}(\bar{T}_{p^d},\mu_{\w})=-\lim_{n\to\infty}\frac{1}{n}\mu_{\w}(\dbrack{\x_1\dots \x_{n}}).
\end{equation}
Thus the connection between two entropies is given by:
\begin{equation}
 H_{\KS}(T_p,\mu_{\w})=\frac{\Gamma}{d}H_{\KS}(\bar{T}_{p^d},\mu_{\w})\label{connection}.
\end{equation}
The coefficient $\Gamma$ is  defined by the limit:
\[\Gamma=\lim_{n\to\infty}m_n/n,\]
where $n$ is the length of the cylinder  $G_n=\dbrack{\varepsilon_0\dots\varepsilon_{n-1} }$ in $\varepsilon$-representation and $m_n$ is the length of the same set $G_n=\dbrack{\x_1\dots \x_{m_n} }$ in the  $\x$-representation. By the Birkhoff's ergodic theorem   this limit is equal  to:
\begin{equation}
 \Gamma=\sum_{i=1}^l q_i \mu_{\w}(I_i).
\end{equation}
The formula for the metric entropy of $H_{\KS}(T_p,\mu_{\w})$ is then follows immediately from (\ref{connection}) and 
the metric entropy (\ref{homogeneousentropy}) of the ``homogeneous'' map $\bar{T}_{p^d}$. Note also that as the right side of (\ref{mainresult})
 amounts to $\Gamma\frac{\log p}{2}$ the proof of the Anantharaman-Nonnenmacher conjecture for the measure 
$\mu_{\w}$ amounts to the proof of 
\[ H_{\KS}(\bar{T}_{p^d},\mu_{\w})\geq\frac{\log p^d}{2}\]
 for the uniformly expanding map $\bar{T}_{p^d}$.


\section{Conclusions and outlook}
In the current paper we  proved  Anantharaman-Nonnenmacher conjecture   for a 
class  of "tensorial" quantizations of one-dimensional piecewise linear  maps $T_p$ whose all slopes are
powers of the  same integer $p$.  It should be stated that we deal here with "tensorial" quantization mostly for the sake of convenience, as
these quantizations allow  very explicit treatment. Actually we believe that the current method with minimal adjustments can be
used to prove the result for all quantizations of maps $T_p$. 
On the other hand, it is clear that the present strategy is restricted to the class of maps
$T_p$, since only these maps can be represented precisely as first return maps for  towers with  uniform expansion rates. To prove the conjecture for  general maps or  Hamiltonian systems (e.g., Anosov geodesic flows)
the current approach must be made more flexible. We believe that such a modification is in fact
possible and it is currently under investigation.  Another question of interest would be about quantum unique ergodicity in quantized one dimensional maps.  Since we know already that various exceptional semiclassical measures appear 
for  the "tensorial" quantizations of maps $T_p$ it would be  interesting to identify an opposite  class of
quantizations for which there are no such sequences at all.

The present application  demonstrates that quantized one-dimensional maps  can be useful as 
toy models for understanding of  general features of quantum chaotic systems. On the technical level 
these systems  are much simpler then Hamiltonian, but still  exhibit generic features 
of chaotic systems. A quite rare opportunity (for chaotic systems) to construct explicit sequences 
of eigenstates make them potentially useful as  test systems.  Another possibility is to use one dimensional maps as models for
scattering systems.  By opening a "gap" in the unite interval 
one can produce quantized
one-dimensional maps with an "absorption" (in  complete analogy with the open
Walsh-Baker maps introduced in \cite{NR}).


\section*{Acknowledgment}

I would like to thank Andreas  Knauf and Christoph  Schumacher for numerous and very helpful discussions on many subjects concerning the paper. 
Most of the  present work was  accomplished during my pleasant stay in Erlangen-Nuremberg University. I am grateful to all my colleagues at the Mathematical Department  for the  hospitality extended to me. 
This work was supported by Minerva Foundation.


\section*{Appendix: Proof of eq.~(\ref{homentropy1})}
Let $T$, $\bar{T}$ be as in Section 7.2 and   $\tilde{T}$ be the corresponding tower map given by (\ref{towermapp}). From the  Markov partition of $I$: $\{\dbrack{\varepsilon_0}, \varepsilon_0\in\{1,2\}\}$  one can easily construct the Markov partition of $\tilde{I}$: $\{\dbrack{\varepsilon_0}\times\{\eta\} , \varepsilon_0\in\{1,2\} \mbox{ and } \eta\in\{0,1\}\}$.  The corresponding n-times refined  (with respect to  $\tilde{T}$) partition is  given then by the set of cylinders: $\{\dbrack{\tilde{\varepsilon}}, \tilde{\varepsilon}=\tilde{\varepsilon}_0\dots\tilde{\varepsilon}_{n-1}\}$, where  $\tilde{\varepsilon}_{i}=(\varepsilon_{i},\eta_i)$,  $\varepsilon_i\in\{1,2\} \mbox{ and } \eta_i\in\{0,1\}\}$.
The metric entropy $H_{\KS}(\tilde{T}, \tilde{\mu})$ is determined  by the corresponding limit of the entropy function:
\begin{equation}
 h_n(\tilde{\mu})=-\sum_{|\tilde{\varepsilon}|=n}\tilde{\mu}(\dbrack{\tilde{\varepsilon}})\log\tilde{\mu}(\dbrack{\tilde{\varepsilon}}).
\end{equation}
For a  cylinder $\dbrack{\tilde{\varepsilon}}$  let 
$\dbrack{\varepsilon}=\pii\dbrack{\tilde{\varepsilon}}$ be the corresponding cylinder in $I$ containing exactly the same sequence of $\varepsilon$ as  in $\tilde{\varepsilon}$. Note that  the time evolution of any point $\tilde{\zeta}\in\tilde{I}$ is completely determined by the sequence $\varepsilon$ and the initial level $\eta_0$. Therefore, for a given $\dbrack{\varepsilon}$  there are precisely two non-empty cylinders $\dbrack{\tilde{\varepsilon}}, \dbrack{\tilde{\varepsilon}'}$ such that $\pii\dbrack{\tilde{\varepsilon}}=\pii\dbrack{\tilde{\varepsilon}'}=\dbrack{\varepsilon}$. 
Furthermore,  $  \tilde{\mu}(\dbrack{\tilde{\varepsilon}})=\Gamma^{-1}\mu(\dbrack{\varepsilon})$, $  \tilde{\mu}(\dbrack{\tilde{\varepsilon}'})=\Gamma^{-1}\mu(\dbrack{1\varepsilon})$ and $h_n(\tilde{\mu})$ can be rewritten as:
\begin{equation*}
 h_n(\tilde{\mu})=-\Gamma^{-1}\sum_{|\varepsilon|=n}\mu(\dbrack{\varepsilon})\log\left({\mu(\dbrack{\varepsilon})}{\Gamma^{-1}}\right)
+\mu(\dbrack{1\varepsilon})\log\left(\mu(\dbrack{1\varepsilon})\Gamma^{-1}\right). \label{tildeentropy}
\end{equation*}
On the other hand, the entropy of the measure $\bar{\mu}$ is given by
\begin{equation*}
 h_n(\bar{\mu})=-\Gamma^{-1}\sum_{|\varepsilon|=n}\left(\bar{\mu}(\dbrack{\varepsilon})+\bar{\mu}(\dbrack{1\varepsilon})\right)\log\left(\frac{\bar{\mu}(\dbrack{\varepsilon})+\bar{\mu}(\dbrack{1\varepsilon})}{\Gamma}\right)
. \label{barentropy}
\end{equation*}
It remains to see that two limits $\lim_{n\to\infty} h_n(\tilde{\mu})/n$, $\lim_{n\to\infty} h_n(\bar{\mu})/n$ coincide. By the convexity of the entropy function
\begin{equation}
 h_n(\bar{\mu})\geq h_n(\tilde{\mu})+\log2 \label{ineqent1}
\end{equation}
Since,  $\log(x+y)\geq\log x$ one also has: 
\begin{equation}
 h_n(\bar{\mu})\leq h_n(\tilde{\mu}).\label{ineqent2}
\end{equation}
From  (\ref{ineqent1}, \ref{ineqent2}) immediately follows the claim.

\end{document}